\documentclass[longauth]{aa} 
 
\usepackage{graphicx}
\usepackage{txfonts}
\usepackage[colorlinks=true, citecolor=blue]{hyperref}
\usepackage{rotating}

\newcommand{\Hmol}{\mbox{H$_{\rm 2}$}}

\newcommand{\kms}{km~s$^{-1}$}
\newcommand{\Msun}{M$_{\odot}$}

\newcommand{\mum}{$\mu$m}

\newcommand{\cc}{\mbox{cm$^{-3}$}}

\usepackage{overpic}
\usepackage{enumitem}

\begin{document}

\title{MICONIC: The multiphase circumnuclear region of Centaurus A as seen with JWST/MIRI MRS observations}
\subtitle{I. Spectral inventory and properties of the warm molecular disk}

\author{L. Evangelista\inst{1}\fnmsep\thanks{E-mail:\url{evangeli@iap.fr}} 
          \and
          P. Guillard\inst{1} 
          \and
          J. Martin \inst{2,3} 
           \and
          P. Salomé\inst{4}
          \and
          A. Alonso Herrero\inst{5}
          \and
          L. Pantoni\inst{6}
          \and
          L. Hermosa Muñoz\inst{5}
          \and
          V. Buiten\inst{3}
          \and
          A. Labiano\inst{7}
          \and 
          M. García-Marín\inst{8}
          \and
          L. Colina\inst{9}
          \and
          T. Böker\inst{8}
          \and
          D. Dicken\inst{10}
          \and
          M.J. Ward\inst{11}
          \and
          G. Wright\inst{10}
          \and
          P. van der Werf\inst{3}
          \and
          S. Garcia-Burillo\inst{12}
          \and
          M. Baes\inst{6}
          \and
          A. Eckart\inst{13,14}
          \and
          G. Östlin\inst{15}
          \and
          D. Rouan\inst{16}
          \and
          F. Walter\inst{17}
         \and
          R. A. Riffel\inst{18,19}  
          \and
          M. Güdel\inst{20,21,22}
}

\institute{
Sorbonne Université, CNRS, UMR 7095, Institut d’Astrophysique de Paris, 98bis bd Arago, 75014 Paris, France
\and
SRON Netherlands Institute for Space Research, Sorbonnelaan 2, 3584 CA Utrecht, The Netherlands
\and
Leiden Observatory, Leiden University, PO Box 9513, 2300 RA Leiden, The Netherlands
   \and
   Observatoire de Paris, PSL University, Sorbonne Université, LUX, 75014 Paris, France
   \and
   Centro de Astrobiología (CAB), CSIC-INTA, Camino Bajo del Castillo s/n, E-28692 Villanueva de la Cañada, Madrid, Spain
   \and
    Department of Physics and Astronomy, Universiteit Gent, Proeftuinstraat 86 N3, B-9000 Ghent, Belgium. 
   \and
   Telespazio UK for the European Space Agency (ESA), ESAC, Camino Bajo del Castillo s/n, 28692 Villanueva de la Cañada, Spain
    \and
    European Space Agency, c/o Space Telescope Science Institute, 3700 San Martin Drive, Baltimore MD 21218, USA
    \and
    Centro de Astrobiología (CAB), CSIC-INTA, Ctra. de Ajalvir km 4, Torrejón de Ardoz, 28850, Madrid, Spain
    \and 
    UK Astronomy Technology Centre, Royal Observatory, Blackford Hill Edinburgh, EH9 3HJ, Scotland, UK
    \and
    Centre for Extragalactic Astronomy, Durham University, South Road, Durham DH1 3LE, UK
    \and
    Observatorio Astronómico Nacional (OAN-IGN)-Observatorio de Madrid, Alfonso XII, 3, 28014 Madrid, Spain
    \and
    Physikalisches Institut der Universität zu Köln, Zülpicher Str. 77, D-50937 Köln, Germany
    \and
    Max-Planck-Institut für Radioastronomie (MPIfR), Auf dem Hügel 69, D-53121 Bonn, Germany
    \and
    Department of Astronomy, Stockholm University, The Oskar Klein Centre, AlbaNova, SE-106 91 Stockholm, Sweden
    \and
    LIRA, Observatoire de Paris, Université PSL, Sorbonne Université, Université Paris Cité, CY Cergy Paris Université, CNRS,92190 Meudon, France
    \and
    Max Planck Institute for Astronomy, Königstuhl 17, 69117 Heidelberg, Germany
    \and
    Departamento de Física, CCNE, Universidade Federal de Santa Maria, Av. Roraima 1000, 97105-900, Santa Maria, RS, Brazil
    \and
    Centro de Astrobiología (CAB), CSIC-INTA, Ctra. de Ajalvir km 4, Torrejón de Ardoz, E-28850, Madrid, Spain
    \and
    Dept. of Astrophysics, University of Vienna, Türkenschanzstr 17, A-1180 Vienna, Austria
    \and
    ETH Zürich, Institute for Particle Physics and Astrophysics, Wolfgang-Pauli-Str. 27, 8093 Zürich, Switzerland
    \and
    ASTRON, Netherlands Institute for Radio Astronomy, Oude Hoogeveensedijk 4, 7991 PD Dwingeloo, The Netherlands
}

\date{Received 01/07/2026; accepted 05/18/2026}

\abstract{

\textbf{Context.}
Supermassive black holes power Active Galactic Nuclei (AGN), injecting energy that may regulate accretion and shapes host galaxies. High–resolution observations of circumnuclear gas and dust are essential to understand these processes.

\textbf{Aims.}
We investigate the morphology, excitation, and kinematics of warm molecular hydrogen (H$_2$) in the inner circumnuclear disk of Centaurus~A, the nearest radio galaxy.

\textbf{Methods.}
We present JWST/MIRI MRS integral-field spectroscopy of the central $170 \times 100$~pc$^2$ at 0.3"–0.7" (5–12~pc) resolution, focusing on pure rotational H$_2$ lines. The spectra exhibit strong nuclear continuum, and bright H$_2$ lines from S(1) to S(8), including the first S(8) detection in Centaurus~A. The lines are optically thin in the nucleus, enabling maps of temperature, column density, and ortho-to-para ratio from spaxel-level excitation diagram fitting.

\textbf{Results.}
Warm H$_2$ shows a complex morphology, dominating the central region where CO emission is weak or undetected. Low-excitation \Hmol\ lines trace an inhomogeneous ring with a 20-pc radius cavity aligned with the jet's near side, suggesting that the jet is affecting the morphology of the molecular disk. Higher-excitation lines form a filamentary structure around the AGN. Kinematics are primarily rotational with an S-shaped distortion, indicating non-circular motions or a warped disk. A coherent, low-dispersion ($\sim$70 km\,s$^{-1}$) streamer spirals inward. A power-law temperature distribution yields a warm (100–2000~K) H$_2$ mass of $(5.6 \pm 1.4)\times10^5$~M$_\odot$ and a dynamical mass of $5\times10^8$~M$_\odot$ within 100~pc. Shock excitation is supported by enhanced H$_2$/continuum and H$_2$/PAH ratios, elevated [NeIII]/[NeII], and sub-equilibrium ortho-to-para ratios (1.6–2.4).

\textbf{Conclusions.}
Turbulent dissipation can balance the observed H$_2$ cooling and likely dominates heating beyond 30~pc. In the inner 100~pc of Centaurus~A, AGN feeding and feedback are linked: shocks excite H$_2$, regulate the gas temperature, and prevent cooling below 100~K, explaining the weak CO emission and lack of a massive outflow. These shocks may drive angular momentum loss and help fuel the nucleus.
}

\keywords{ galaxies: evolution -- galaxies: active -- galaxies: nuclei -- 
            galaxies: jets -- galaxies: ISM -- 
            galaxies: individual: Centaurus A
               }

\titlerunning{The molecular circumnuclear disk of Centaurus A with JWST}
\maketitle
\nolinenumbers
%

\section{Introduction}\label{sec:intro}

The feeding and feedback of supermassive black holes (SMBH) represent intricate processes that shape the dynamics and evolution of galaxies. Matter accretes onto the black hole powering an Active Galactic Nucleus \citep[AGN, e.g.][]{capelo_black_2023}. The released radiative and mechanical energy generates powerful feedback mechanisms such as gas heating and jet-driven winds, that can influence the interstellar medium of its host galaxy and ultimately regulate its star formation \citep[see][for a review]{veilleux_cool_2020}. 
Even though the fueling mechanisms are not fully understood, merger events and disk instabilities in the spirals and bars are thought to trigger the inflow of matter to the nucleus \citep[e.g.][]{Combes2001}. A corona of hot material forms above the accretion disk and can inverse-Compton scatter photons up to X-ray energies \citep{Haardt_91}. The radiation from the accretion disk excites cold atomic material close to the black hole and this radiates via emission lines. A large fraction of AGN output can be obscured by the interstellar medium \citep[ISM; i.e. dust and gas; ][]{lusso_2013} close to the accretion disk, which in turn re-radiates in the infrared. Investigating the accretion and AGN-driven outflow triggering mechanisms requires high spatial and spectral resolution observations close to the jet launching site to enable detailed morpho-kinematical studies of the gas and dust in the circum-nuclear regions.    

At a distance of (3.8$\pm$0.1)~Mpc \citep{harris_2010}, the peculiar elliptical galaxy Centaurus A (NGC~5128, hereafter Cen~A) is the nearest radio galaxy, thus offering a unique opportunity to study AGN environments at parsec scales in the IR band. It is a prototype Fanaroff-Riley Class I \citep[low luminosity,][]{fanaroff_1974} radio galaxy \citep{israel_centaurus_1998}, and as such it is an ideal laboratory to understand a major class of active galaxies, and associated AGN feedback processes. It hosts a SMBH with a mass between 4.5--5.5 $\times 10^7$~\Msun, based on molecular hydrogen (\Hmol) and stellar kinematics \citep{neumayer_central_2007}. 

\begin{figure*}[!t]
    \centering
    \includegraphics[width=0.9\linewidth]{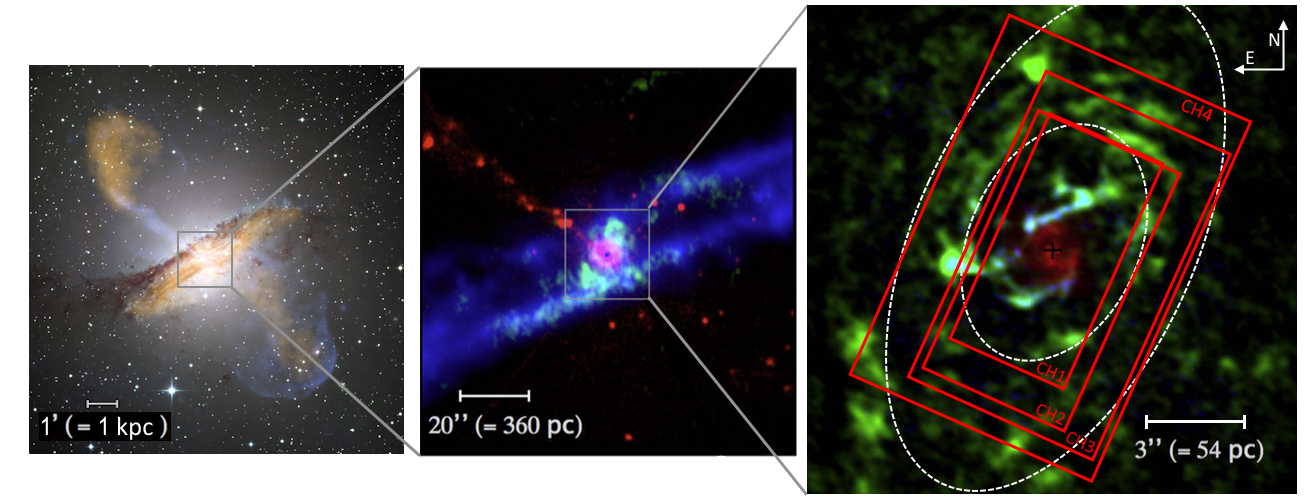}
    \caption{Zoom into the inner region of Centaurus~A adapted from \citet{espada_disentangling_2017}. \textit{Left:} Color composite image of Centaurus A. Credit: ESO/WFI - Optical; MPIfR/ESO/APEX/\citet{Wei__2008} - Submillimeter; NASA/CXC/CfA/\citet{KRAFT2003} - X-ray. \textit{Center:} integrated CO(2-1) emission map from SMA (green) \citep{espada_2009}; dust emission at 8~\mum\ from Spitzer/IRAC (blue) \citep{quillen_spitzer_2006}; the jet in X-ray from Chandra (red) \citep{KRAFT2003}. \textit{Right:} integrated CO(3-2) and CO(6-5) maps from ALMA (respectively green and blue); \Hmol\,1--0 S(1) map at 2.122~\mum\ from VLT/SINFONI (red). The red rectangles in the right panel are the mosaic footprints of the four MRS channels. The dashed white ellipses outline the CND as defined by \citet{espada_disentangling_2017}.  }
    \label{fig:MRSfootprints}
\end{figure*}

The morphology of Cen~A has been thoroughly mapped in existing literature, from kiloparsec to parsec scales. It features powerful jets (PA$_\mathrm{jet}=51^\circ$) with lobes that extend 250~kpc out of the AGN \citep{israel_centaurus_1998}. The galaxy itself contains an outer star-forming molecular disk of 4~kpc in diameter \citep{espada_star_2019}, and exhibits a prominent dust lane of 3~kpc in diameter \citep{quillen_spitzer_2006}. The disk has a nearly edge-on warped geometry \citep{quillen_warped_1993,quillen_warped_2010}.  Mid-infrared imaging ot the dust lane with \textit{Spitzer}/IRS reveals emission from S(0), S(2), S(3), and S(5) rotational transition lines of \Hmol\ \citep{quillen_2008}. \textit{Spitzer}/IRAC and MIPS reveal a bright central unresolved source as well as a "parallelogram–like" emitting structure across the region covered by the central dust lane \citep{quillen_spitzer_2006}. This parallelogram structure has also been observed in the mid-IR by ISOCAM, in the sub-mm by SCUBA, and is understood to be the remnant of a minor merging event \citep{israel_centaurus_1998}. The center of the galaxy hosts a circumnuclear disk (CND) of molecular gas, mapped for the first time by \citet{espada_2009}. It is described by \citet{espada_disentangling_2017} as an ellipse with major and minor axes of, respectively, 20"$\times$10", or 360~pc$\times$180~pc, and PA$_\mathrm{major}=155^\circ$ (without accounting for projection effects). The overall morphology of Cen~A as described in this paragraph is depicted through different scales in Figure \ref{fig:MRSfootprints}.

ALMA CO(3–2) line observations in the inner $\sim$400~pc of the CND (Fig.~\ref{fig:MRSfootprints}, right panel, green) reveal a clumpy distribution with large cavities, while the CO(6–5) emission (Fig.~\ref{fig:MRSfootprints}, right panel, blue) is restricted to filaments located north and south of the nucleus \citep{espada_disentangling_2017}. 
In the $\sim100$~pc Innermost region of the CND (ICND) (see Fig.~\ref{fig:diag}), \textit{Spitzer}/IRS spectroscopy reveals strong \Hmol\ rotational line emission, from S(0) to S(7) \citep{Ogle_2010}, with an estimated mass of warm (in a temperature range of 100-1000K) \Hmol\ gas of $3.6\times 10^7$~\Msun, in the central 3.7" $\times$ 3.7". \citet{Ogle_2010} classify Cen~A as a molecular hydrogen emission galaxy (MOHEG), based on its ratio $\mathrm{Log}_{10}\left(L_\mathrm{H_2}/L_{\mathrm{PAH}_{7.7\mu m}}\right)=-0.8\pm0.1$ exceeding what is expected from models of UV and X-ray heating of the molecular gas \citep{Guillard_2012}. This indicates that additional excitation processes, like shocks, are required to account for the elevated $L_{\Hmol}/L_{\mathrm{PAH}_{7.7\mu m}}$ values.
However, the limited spectral and spatial resolution of IRS prevented any determination of the spatial distribution and kinematics of the mid-IR pure rotational \Hmol\ lines. Moreover, VLT SINFONI observations reveal the presence of a nuclear disk (ND) of hot molecular gas $\sim50$~pc in diameter (Fig.~\ref{fig:MRSfootprints}, right panel, red), as traced by the \Hmol\ 1–0 S(1) ro-vibrational line emission \citep{neumayer_nucleus_2007}.

In this paper, we use data from the James Webb Space Telescope's Mid-Infrared instrument (\textit{JWST}/MIRI) medium resolution spectrograph \citep[MRS, ][]{Wells_2015, argyriou_jwst_2023, wright_2023} of Cen~A which were obtained as part of the MIRI GTO program "Mid-Infrared Characterization of Nearby Iconic galaxy Centers" (MICONIC) of the MIRI European Consortium. This joint MIRI + NIRSpec GTO program aims to observe the central region of Cen~A to obtain a detailed study of the influence of the SMBH on its environment.

Using JWST/MRS observations of Cen~A, we probe the distribution and kinematics of the molecular and ionized gas in the inner $\sim100$~pc region of the CND,  with a resolution between 0.3" and 0.7" (5 to 12 pc), depending on the wavelength: a 50 times greater sensitivity and 7 times higher angular resolution than \textit{Spitzer}. 

This paper is the first of two focusing on the \Hmol\ emission in the CND, and focuses on presenting the observational results, while Paper~II will present the modelling of the \Hmol\ excitation. Two other companion papers on MRS observations of Cen~A are presenting results focused on the ionized gas kinematics \citep{alonso-herrero_miconic_2025} and the properties of the polycyclic aromatic hydrocarbons (PAH) emission  \citep{pantoni_2026}.
The MICONIC program also includes other targets like NGC 6240 \citep{munoz_2025}, Mrk 231 \citep{alonsoherrero2024}, Arp 220 \citep{buiten_2025}, and the region surrounding SgrA$^\star$.

The paper is outlined as follows. In Sect. 2 we describe the JWST/MIRI MRS observations as well as our data reduction. In Sect.~3 we present a spectral line inventory, our spectral maps, \Hmol\ excitation, and mass estimates. Sect.~4. discusses qualitatively the \Hmol\ excitation (models will be presented in paper~II) and the comparison between the \Hmol\ and ionized gas \citep[the kinematics is further discussed in][]{alonso-herrero_miconic_2025}.

\section{JWST MIRI MRS observations and data analysis}

\subsection{Observation strategy and data reduction}

This work uses data obtained during the JWST Cycle 1 Guaranteed Time Observations (GTO) program PID 1269, performed on March 19, 2023. The MRS spectra cover a total wavelength range from $4.9 - 27.9\mu m$, separated in four integral field units (IFUs) referred to as channels (hereafter CH1, 2, 3, and 4), each split in three  bands (short, medium, long, hereafter A, B, C). The channels cover slightly different fields of view (FoV), from 3.6\arcsec$\times$7.5\arcsec\ (CH1) up to 7.7\arcsec$\times$12\arcsec\ (CH4), and have different spatial (from 0.19\arcsec to 1\arcsec) and spectral (from $\sim$3700 to $\sim$1500) resolutions \citep{labiano_wavelength_2021}. Spatial and spectral samplings for each channel and sub-channel are provided in Appendix~\ref{appendix:mrs}, along with the mean spatial and spectral resolutions \citep{law_3d_2023,labiano_wavelength_2021,pontoppidan_2024}, as well as the centering and size of the respective FoVs. The P.A. of the FoV is $\sim-67^\circ$.
 We used a $2\times1$ mosaic of the central region of Cen~A. The footprints of the FoV of the MRS channels are shown in Figure~\ref{fig:MRSfootprints}. We used the four-point extended source dither pattern, which helps to minimize the continuum wiggles due to the undersampling of the point spread function (PSF) \citep{law_3d_2023}. We set 10 groups per integration, and 5 integrations per exposure, in FASTR1 readout mode, covering the whole MRS spectral range in three exposures (one per MRS band), for an on-source integration time of 600s per MRS band. Using the recommended strategy, we took a free-of-source background observation to the West of the galaxy group with the two-point extended source dither pattern and the same integration time per band.

The data reduction was done with the JWST Science Calibration Pipeline \citep[version 1.15.6,][]{bushouse_jwst_2023}, with the context 1293 for the Calibration References Data System following the standard procedures \citep[see e.g.,][]{labiano_miri_2016, alvarez-marquez_nuclear_2023}, for detailed examples of MRS data reduction and calibration). All of the individual raw images were first processed for detector-level corrections using the \textit{Detector1Pipeline} module of the pipeline.
Careful examination of the data showed that the stage 1 corrections \citep{morrison_jwst_2023} could be run with default parameters, except the cosmic-ray flagging jump rejection threshold that we lowered to 3.5$\sigma$, and the cosmic ray shower flagging that we switched on. 

Taking advantage of the off-source background images, we used the image-to-image (2D pixel-by-pixel) background correction in the stage 2 pipeline 
\citep[Spec2, ][]{argyriou_jwst_2023, gasman_jwst_2023, patapis_geometric_2024}, and the 2D residual fringe correction (removing fixed-frequency modulations in the spectrum caused by standing waves). To remove hot/warm pixels, we flagged outliers in the background exposures and masked them in both the background and science frames by turning on bad pixel self-calibration in the Spec2 pipeline which uses all dithered exposures of a given detector to find and flag bad pixels that may have been missed by the bad pixel mask. We set the maximum fraction of pixels to flag to 0.5\%. 
We switched off the sky matching step in the stage 3 of the pipeline, before producing the final fully reconstructed science cubes \citep{law_3d_2023} as they introduced artifacts in the data\footnote{This is acceptable given the strong brightness of the target over the whole FoV.}. 
Since the MRS has a small field of view, its absolute astrometric solution cannot always be tied to an external reference frame using MRS data alone, so we used simultaneous imaging to improve the astrometric solution of the MRS. We re-aligned the MRS astrometry thanks to the simultaneous imaging data registered to the GAIA DR3 catalog, resulting in less than 0.1" residuals. Finally, the science cubes were rotated to the usual orientation with north up and east to the left, resulting in FoV sizes ranging from 7.5\arcsec$\times$3.8\arcsec for CH1 to 11.7\arcsec$\times$7.2\arcsec for CH4.

\subsection{Analysis of the spectral cubes}\label{sect:analysis_cubes}

For each of the twelve spectral cubes, we performed individual 1D-spectral extractions over the respective full FoV. We did not perform additional 1D de-fringing because we extracted the nuclear spectrum over an aperture 2 times larger than the PSF size. Thanks to the improved spectrophotometric calibration \citep{gasman_jwst_2023, law_2025}, the extracted spectra did not require stitching of bands, except a small scaling factor (0.97) to align subchannels CH1A and CH1B, the latter assumed to be better calibrated than the former due to its higher signal-to-noise ratio (SNR) \citep[see ][]{law_2025}.
In what follows, we detail our workflow for the analysis of the spectral cubes.

\textit{Line identification:} The emission lines were identified using the line list available on the ISO Spectrometer Data Center website\footnote{\url{https://www.mpe.mpg.de/ir/ISO/linelists/index.html}}, correcting for the expected redshift of CenA $z=0.001825$ \citep{salome_star_2016}.

\textit{Preliminary 1D line fit:} We fit the identified lines in the spectra averaged over the full FoV. As a first approximation, each line is fitted with a Gaussian profile to estimate the spectral full width at half maximum (FWHM$_\lambda$) and locate the center of the line. The spectral extent of the line is identified with a $3\sigma_\lambda$ criterion around the center of the line, where we define $\sigma_\lambda=\mathrm{FWHM_\lambda}/(2\sqrt{2\,\mathrm{ln2}})$.  

\textit{Continuum subtracted sub-cubes:} To ease handling of the cubes, we extract a set of spectral slabs (sub-cubes) around each line. For each sub-cube, we perform a spaxel-by-spaxel linear fit of the continuum along the spectral axis. The spectral range of these sub-cubes is narrow enough to allow a low residual linear fit of the continuum. The fitted continuum is then subtracted. The sub-cubes thus obtained are used to generate moment maps and to calculate line fluxes. In Appendix~\ref{appendix:spectra}, Fig.~\ref{fig:lineprof} shows the line profiles extracted from the sub-cubes. The baseline levels illustrate that the continuum removal is satisfactory. 

\textit{Continuum maps:} We extract a second set of sub-cubes, that are not continuum-subtracted, to generate maps of the continuum. For each sub-cube, we mask the line and average the remaining continuum spaxel-by-spaxel along the spectral axis. To obtain a continuum map (in W m$^{-2}$ sr$^{-1}$), we then integrate the mean continuum level over spectral intervals equivalent to the extent of the adjacent line (integration bounds marked in black on Fig.~\ref{fig:lineprof}). The continuum maps are used to identify the location of the AGN, via 2D-Gaussian fit of the bright centroid (see Sect.~\ref{subsec:continuum} and App.~\ref{appendix:maps} Fig.~\ref{fig:continuum}). We hence adopt the measured AGN RA-Dec angular position \hbox{13:25:27.63 -43:01:08.30} (J2000).

\textit{Masking the central spaxels:} The central spaxels present a continuum level up to 2 orders of magnitude stronger compared to the ICND. In case of high-contrast signals, the non-linearity, charge migration, and the scattering of photons in the MIRI detectors have 2 effects on the PSF: a broadening effect (similar to the brighter-fatter effect seen in CCDs) and spectral fringes \citep{law_3d_2023, argyriou_2023c}. The intensity of the fringes reaches the same order of magnitude as the intensity of the lines for the spaxels within 0.23" (4.24 pc), 0.3" (5.53 pc), 0.4" (7.37 pc), and 0.7" (12.89 pc) from the AGN, respectively for CH1, 2, 3, and 4, which makes line detection difficult in the ND. Since the corrections of these effects are still limited in the current JWST pipeline \citep{gasman_2025}, we have conservatively masked those central spaxels where fringes could affect both the line fluxes and kinematics. Outside of this mask, we carefully inspected the baselines and potential velocity shifts in the side lobes. At the wavelengths of the low-J lines, 10, 12 and 17~\mum, the PSF broadening effect on the FWHM is small \citep[5, 3 and 2\%,][]{gasman_2024}, and we see no effect on the moment-1 maps. The continuum maps are used to measure the angular size of the central continuum-dominated region. We perform a 2D-Gaussian fit of the nuclear bright area and generate a circular mask of radius 3$\sigma_\mathrm{nuc}$ (where $\sigma_\mathrm{nuc}$ is the 2D-Gaussian fitted radial extent of the area). The mask is then applied to the continuum subtracted sub-cubes to exclude the nuclear fringe-dominated spaxels. 

\textit{Line detection and spaxel flagging:} For each continuum-subtracted spaxel, the continuum standard deviation $\sigma_\mathrm{STD}$ defines the noise level, and a line is considered detected if its peak intensity exceeds this noise by a threshold of 3$\sigma_\mathrm{STD}$. Spaxels not meeting these conditions are attributed a NaN value.  For the line \Hmol\ 0--0 S(8), that exhibits a lower SNR, we tested both a 2$\sigma_\mathrm{STD}$ and a 1$\sigma_\mathrm{STD}$ criterion. 

\textit{Moment map extraction:} Moment-0 maps are constructed by integrating the continuum-subtracted sub-cube over the spectral extent of the line, while velocity and velocity dispersion maps are obtained from a spaxel-by-spaxel 1D-Gaussian fit to the line profile. Velocity dispersion maps are corrected for MRS spectral resolution. This is performed via quadratic subtraction of the observed line width $\sigma_\lambda$ with the expected spectral resolution width at the specific wavelength, as characterized by \citet{pontoppidan_2024}.

\textit{Map convolution and re-projection:} 
In order to generate maps of the \Hmol\ physical parameters (e.g. temperature, column density) from the observations, we need to fit \Hmol\ excitation diagrams on a spaxel-by-spaxel basis. This requires maps from different MRS channels to be convolved to the same resolution and projected on a common spatial grid. The convolution is performed with a Gaussian kernel, with FWHM set to match the resolution of the least-resolved line (0.67" or 12~pc for the \Hmol\ 0--0 S(1)) and amplitude normalized to preserve flux. For the re-projection, we use Python's \texttt{reproject\_interp} routine to preserve flux and choose the intermediary grid of CH2 with a spaxel size of 0.17" (or 3~pc).

\section{Results}

\subsection{Centaurus A mid-IR spectral inventory}
\label{sec:inventory}

\begin{figure*}[ht]
\hspace{-0.5cm}
       \begin{overpic}[width=1.0\linewidth]{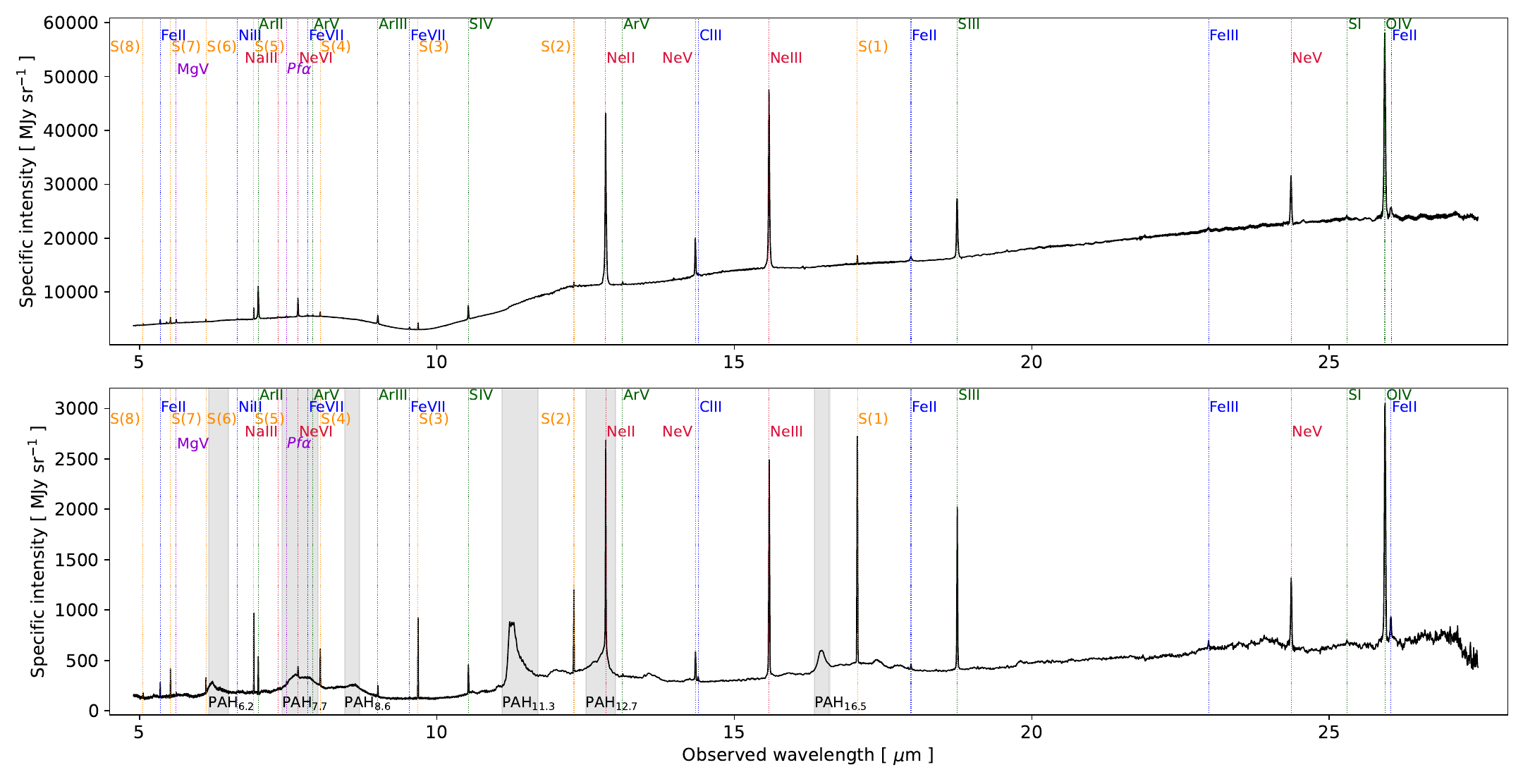}

        \put(65, 20){
            \includegraphics[width=0.12\linewidth, trim=115 35 95 40, clip]{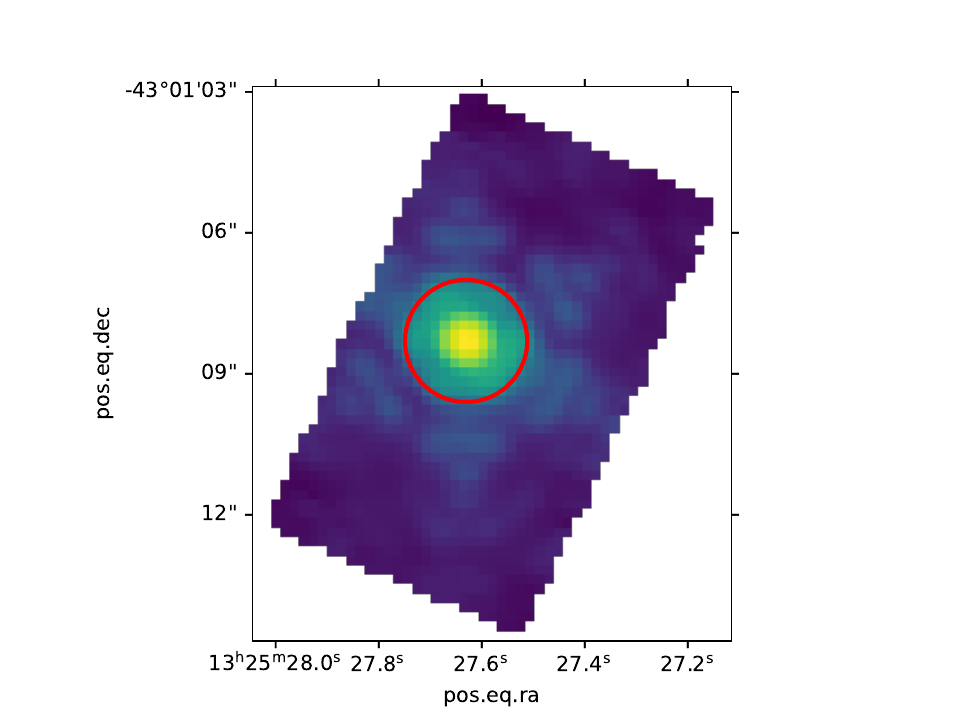}}

        \put(70.75, 28){\color{red}{\vector(0,2){6}}} 
        \put(72, 24){\color{red}{\vector(0,-2){14}}}        
    \end{overpic}

\caption{Nuclear (top) and circumnuclear (bottom) averaged spectra obtained from the four channels of MIRI-MRS. The two regions of extraction are delimited by a 1.3"-radius circle (24~pc), as shown on the small inset image (continuum map at 17~\mum). This aperture corresponds to $2\times$FWHM of the PSF at the wavelength of 0--0 S(1) line. The identified emission lines are labeled in different colors. Brackets are omitted from the spectroscopic notation for visual clarity. The \Hmol\ rotational lines are labeled in yellow. The main PAH features are indicated with gray vertical bands. Zooms of the spectra on the different MRS channels are displayed in Appendix~\ref{appendix:spectra}, Figure~\ref{fig:spectra}.}
\label{fig:spectrum}
\end{figure*}

The spectra extracted from the JWST/MRS data cubes are shown in Figure~\ref{fig:spectrum}, averaged over two regions of the FoV\footnote{The spectra from the different channels are shown in Appendix~\ref{appendix:spectra} (Figure~\ref{fig:spectra}), averaged over the same regions.}: an inner region, located inside a 1.3"-radius (24 pc) from the AGN, covering the ND, and a surrounding region covering the ICND. We chose 
an extraction aperture radius about 2 times larger than the PSF at $17$~\mum. The spectra show a rich collection of emission lines, including both molecular hydrogen and recombination lines. In particular, we identify the suite of pure rotational lines of \Hmol\ from S(1) to S(8), the latter being observed for the first time in Cen~A. Table~\ref{table:lines_data} details the wavelengths and widths of the \Hmol\ purely rotational lines. We also identify emission lines from the ionized gas, such as [SIII] at 18.75~\mum, [NeIII] at 15.583~\mum, [NeII] at 12.837~\mum, and [ArII] at 6.98~\mum. We redirect the reader to \cite{alonso-herrero_miconic_2025} for a detailed analysis of the ionized gas. Finally, we observe signatures of PAHs in the ICND. A detailed analysis of the PAH features is performed by \citet{pantoni_2026}. 

Line fluxes are listed in Appendix~\ref{appendix:spectra} (Table~\ref{tab:flux}) for different regions of the FoV. As the lines are spectrally resolved their fluxes are calculated by integrating the continuum-subtracted profiles (shown in Figure~\ref{fig:lineprof} with respective integration boundaries). We applied the aperture correction factors provided by \citet{law_2025} (11--13\% for the nucleus and 5\% for the larger apertures). For ionized gas lines that have non single-Gaussian profiles (e.g. asymmetrical or a broad underlying velocity component, like [NeII] or [ArII]), we find nuclear fluxes 10--15\% larger than those computed by single Gaussian fitting \citep{alonso-herrero_miconic_2025}, because we account for the underlying broad velocity component.

\subsection{Spatial distribution of the molecular gas}\label{sect:morph}

\begin{figure*}[ht]
    \centering
        \includegraphics[width=0.8\textwidth, trim=0 20 0 8, clip]{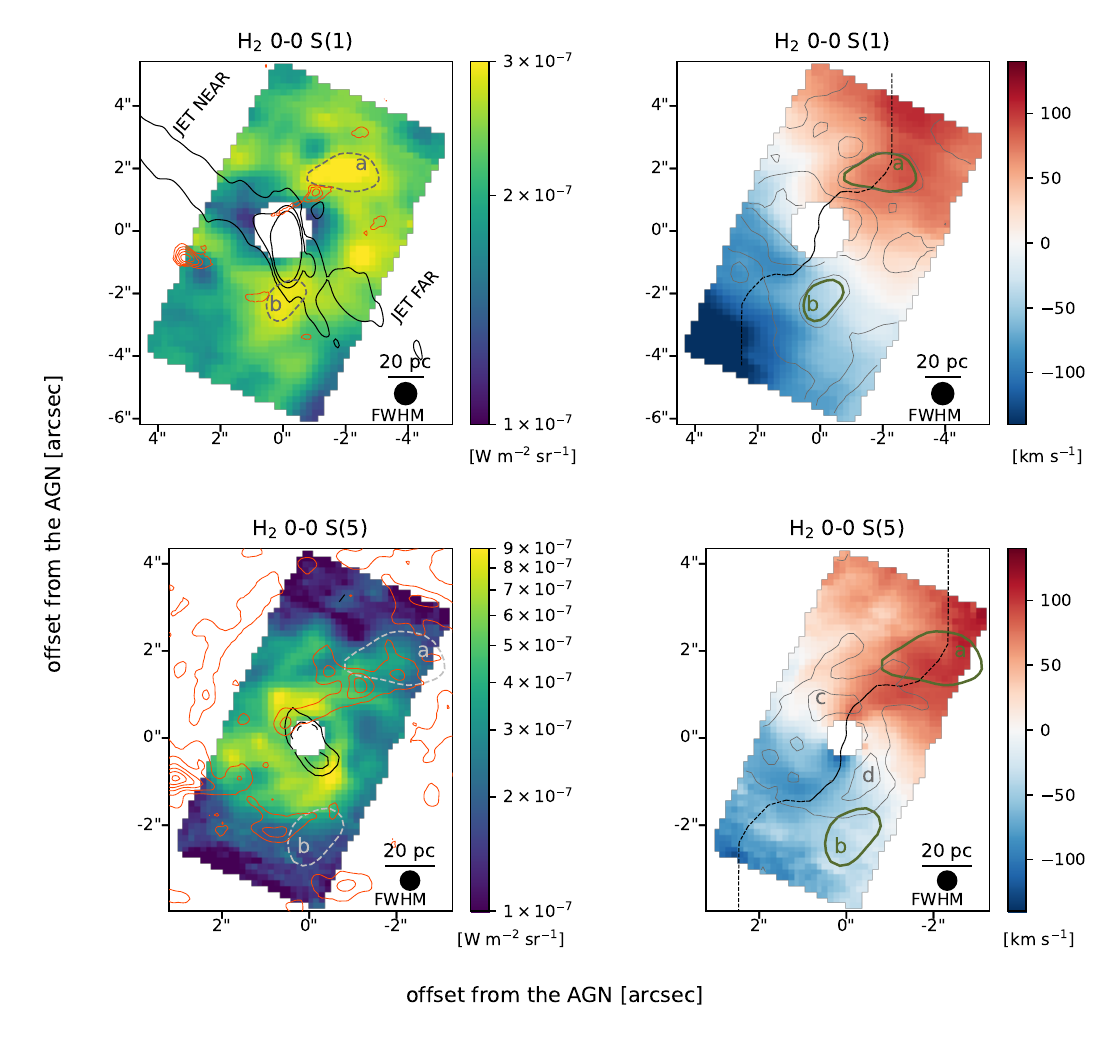}
       \caption{Surface brightness maps (left) and velocity maps (right) of the \Hmol\ lines 0--0 S(1) at 17~\mum\, and S(5) at 6.9~\mum\, with central spaxels masked due to spectral fringing (see Sect.~\ref{sect:analysis_cubes}). The FWHM of the MRS PSF of the respective channel is shown in the lower right corner. The black contours on the top left map are 8.5~GHz radio VLA contours  (0.22, 3.3, 16~mJy~beam$^{-1}$) from \citet{hardcastle_radio_2003}, tracing the jet that aligns with a central cavity seen on the S(1) (note that the VLA beam has a strong North-South elongation). The orange contours on the top left map are 434~\mum\ ALMA CO(6--5) contours (0.01, 0.07, 0.12, 0.18, 0.24, 0.30~Jy~km~s$^{-1}$~beam$^{-1}$), while those on the bottom left panel are 870~\mum\, ALMA CO(3--2) contours (-0.10, 1.17, 2.45, 3.72, 5.00, 6.28~Jy~km~s$^{-1}$~beam$^{-1}$) from \citet{espada_disentangling_2017}. CO(6-5) is scarce, and most of CO(3--2) shows large cavities filled by warm \Hmol\ emission. The black contours on the bottom left map are JWST infrared contours tracing [Ne{\sc \ vi}] at 10.51 \mum, which features an alignment with the \Hmol\ 0--0~S(5) hot patches and with the jet. The gray dashed contours on the left maps highlight the two bright hotspots (a) and (b) visible on the S(1) map. The hotspots align with the filaments on the S(5) map and the northern filament visible in CO(3--2) and CO(6--5). The same contours are superimposed in green on the right-side maps, along with the gray contours tracing the relative surface brightness maps, and the dashed black line tracing the line of the nodes of the warped-disk model from \citet{neumayer_central_2007}. The contours of the hot S(5) patches aligned with the jet <20pc from the AGN are labeled (c) and (d) on the bottom right panel.}
    \label{fig:moment 0 maps}
\end{figure*}

\begin{figure}[ht]
    \centering
    
        \includegraphics[width=0.8\linewidth]{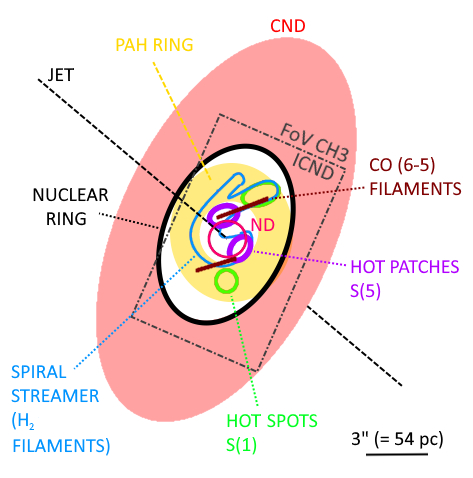}
    
    \caption{Sub-kpc scale schematic of the center of the Cen~A. The dotted straight line represents the direction of the jet. The red semitransparent annulus represents the molecular CND. The black ellipse represents the nuclear ring of CO described by \citet{espada_disentangling_2017}. The brown bars north and south of the AGN represent the filaments of CO(6-5). The blue shape traces the contours of the low dispersion spiral of warm \Hmol\ filaments (see Fig.~\ref{fig:moment 0 maps} and Fig.~\ref{fig:s5m2}). The green ellipses represent the S(1) hot spots (a) and (b) (see Fig.~\ref{fig:moment 0 maps}, upper left). The purple ellipses represent the S(5) hot patches (c) and (d) (see Fig.~\ref{fig:moment 0 maps}, bottom left). The pink inner empty ring in the center represents the ND of hot gas analyzed by \citet{neumayer_central_2007}. For reference, the dashed-dotted rectangle represents the FoV of CH3. The ICND is the region of the CND covered by our FoV (including \Hmol). The yellow semitransparent annulus represents the PAH ring analyzed by \citet{pantoni_2026}}.
    \label{fig:diag}
\end{figure}

\begin{figure}[h]
    \centering
    \includegraphics[width=0.90\linewidth, trim=0 0 0 0, clip]{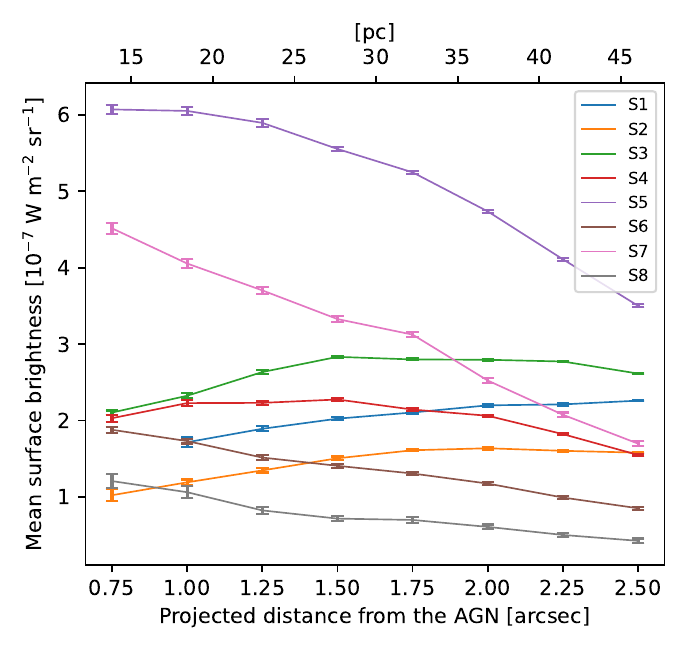}
    \caption{Radial profile of the mean surface brightness as a function of the projected distance from the AGN. Brightness has been averaged within concentric elliptical annuli, of ellipticity 0.5 and PA=155$^\circ$ (following the geometry of the CND). The thickness at the major axis is 0.25" (4.6~pc) for each annulus. The S(1) and S(2) lines grow brighter at a larger distance from the AGN, while lines from the S(5) to S(8) peak near the AGN and grow fainter farther out.}

    \label{fig:radial profil flux}
\end{figure}

The surface brightness maps of the molecular hydrogen lines 0--0 S(1) and S(5) are presented in Figure~\ref{fig:moment 0 maps} (left-side panels), which illustrates the heterogeneous morphology of the gas at different excitation levels\footnote{The surface brightness maps of the other \Hmol\ lines can be found in Appendix~\ref{appendix:maps} (Figure~\ref{fig:M0})}. The maps show diffuse emission, with clumps and filaments spreading over distances of the order of 10~pc, forming a complex disk-like structure. Low excitation \Hmol\ is brighter in the ICND, whereas high excitation \Hmol\ peaks near the ND. 

The S(1) map presents two bright hotspots, one north (a) and one south (b) of the AGN. The ND presents a cavity that overlaps the superimposed VLA radio contours of the jet in black \citep{hardcastle_radio_2003}. This alignment suggests that the jet interacts directly with the weakly excited \Hmol.

The S(5) map shows a morphology that is opposite to that of the S(1). The central cavity disappears in favor of two hot patches, (c) and (d), extending for 1" (18~pc), respectively North-East and South-West of the AGN (see contour levels in the bottom right panel of Fig.~\ref{fig:moment 0 maps}). These patches align with the black contours of the [Ne{\sc \ vi}] ionized gas line at 10.51~\mum\, which shows morphological correlation with the jet \citep{alonso-herrero_miconic_2025}. The white contours locate the hotspots (a) and (b) of the S(1) map, revealing alignment with the S(5) filaments, especially to the north. The filaments appear to gather into what we define as a spiral streamer (see Sect.~\ref{subsec:H2kinematics}).

Figure~\ref{fig:diag} provides a schematic of the morphology of the main molecular gas features, consistently with existing literature. The figure also shows the PAH ring studied by \citet{pantoni_2026} which displays the same disky structure of the S(1), with a distinct PAH-deficient area overlapping the (a) hotspot.

The orange contours from ALMA trace the lines of CO(6-5) at 434~\mum, on the S(1) panel, and CO(3-2) at 870~\mum, on the S(5) panel. The contours show the relative scarcity of cold gas in the FoV, with CO(3-2) presenting large cavities filled by \Hmol. Another possible interpretation of the CO cavities may be flux loss due to low spatial frequency filtering in the SMA and ALMA interferometers. 

The CO(6-5) contours show two filaments, respectively north and south of the AGN  that intersect the \Hmol\ filaments at the location of the hotspots (a) and (b). These bright intersection points have already been presented by \cite{espada_disentangling_2017} who describe their emission as shock-driven. CO(6-5) also presents a clump to the east (RA $\sim3$\arcsec\ from the AGN) that overlaps a faint cavity in \Hmol.  

Figure~\ref{fig:radial profil flux} plots the surface brightness radial profiles of the \Hmol\ lines showing the radial gradient of the excitation. \Hmol\ emission peaks in the outer regions of the map for the S(1) and S(2), halfway through for the S(3) and S(4), and near the center for all the other lines from S(5) to S(8).

The masked spaxels at the center of the maps conceal the innermost part of the region analyzed by \citet{neumayer_central_2007} where SINFONI observations of the \Hmol\ 1--0 S(1) line at 2.122~\mum\ reveal emissions of hot molecular gas within 3" (54~pc) from the AGN. Most of the warm \Hmol\ in the ICND is hence at the interface between the hot molecular gas of the ND and the cold CO of the CND. This result, along with the radial gradient in the peaking excitation for \Hmol, is consistent with the gradient in the CO(6-5)-to-CO(3-2) flux density ratio reported by \citet{espada_disentangling_2017}.

\subsection{Molecular gas kinematics}
\label{subsec:H2kinematics}

The velocity maps for the \Hmol\ 0–0 S(1) and S(5) lines\footnote{Velocity and dispersion maps for all \Hmol\ transitions are shown in Appendix~\ref{appendix:maps} (Figures \ref{fig:M1} and \ref{fig:M2}).} are presented in Figure~\ref{fig:moment 0 maps} (right-side panels). They reveal a global rotation pattern around an axis at PA$_\mathrm{rot}\sim50^\circ$, which is similar to PA$_\mathrm{jet}$ measured by \citet{neumayer_central_2007}. The line-of-sight velocities range from $[-180, +110]$~\kms\ for the S(1)\footnote{The skewness in the S(1) velocity field is localized to a narrow, blueshifted region in the southeastern corner of the map, at the limit of the FoV.} and $[ -120, +120]$~\kms\ for the S(5). 
For the ionized gas lines such as [NeII] at 12.837~\mum, [ArII] at 7~\mum, and [FeII] at 5.3\mum, \citet{alonso-herrero_miconic_2025} found significant velocity dispersions up to 3 times stronger than \Hmol. Table~\ref{table:lines_data} provides the velocity dispersions for the \Hmol\ lines, obtained via a single Gaussian fit, in the full FoV and in the different sub-regions. The velocity dispersion is corrected for MRS spectral resolution.

We observe non-circular motion components, most notably a characteristic S-shaped distortion, visible in white (close to the systemic velocity) in all \Hmol\ velocity maps, already noted by \citet{alonso-herrero_miconic_2025}. The overlaid contours in Figure~\ref{fig:moment 0 maps} trace the level lines from the brightness maps in gray, the (a) and (b) hotspots in green, and the line of the nodes of the warped disk model from \citet{neumayer_central_2007} in black. If the hotspots are the result of higher column density regions due to projection effects on a warped-disk model, we would expect them to align with the line of the nodes. The (a) hotspot from the S(1) map aligns with the redshifted segment of the S-shaped distortion and follows the line of the nodes, however the (b) hotspot does not show the same alignment. \citet{alonso-herrero_miconic_2025} compare the observed velocity field with a $^\mathrm{3D}$BAROLO warped-disk model \citep{Teodoro_2015} and find residuals between $-$40 and 40 \kms, proving that the warp alone is insufficient to explain the distortion in the velocity field. Therefore, non-circular components must also be present as already concluded by these authors. 

\begin{figure}[t]
    \centering
    \includegraphics[width=0.75\linewidth, trim = 0 0 0 0 , clip]{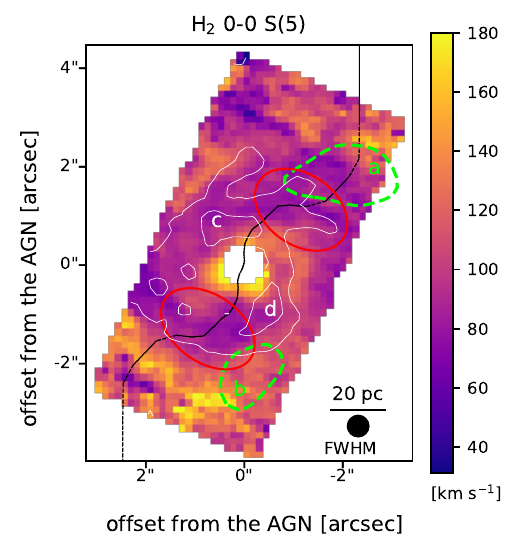}
    \caption{Velocity dispersion map of the \Hmol\ 0--0 S(5) line. Green contours trace the (a) and (b) hotspots identified in the S(1) map. The white contours represent the surface brightness of the line, with the (c) and (d) hot patches labeled. The black line indicates the line of the nodes of the warped disk model from \citet{neumayer_central_2007}. The low-dispersion (70-90~\kms) spiral streamer overlays the S(5) filaments (white contours) and is consistent with a coherent bulk gas flow. The red ellipses trace the regions where 1D-Gaussian fits of the S(1) and the S(5) reveal residuals $>3\sigma_\mathrm{STD}$ (see Fig.~\ref{fig:disp_profiles}). }
    \label{fig:s5m2}
\end{figure}

Figure~\ref{fig:s5m2} shows the velocity dispersion map of the \Hmol\ 0--0 S(5) line, which presents a low velocity dispersion (FWHM$\sim$70-90~\kms) spiral structure, 20~pc wide, that encompasses the \Hmol\ filaments (Fig.~\ref{fig:moment 0 maps}, left). This coherent low dispersion structure, along with the homogeneous surface brightness and \Hmol\ excitation (see Sect.~\ref{subsec:H2excitation}) are features commonly observed in systems where a bulk inflow has been identified \citep{Mueller_2009,Dominguez_2020,Tanaka_2026}. For such reasons, we refer to this structure as a spiral streamer.

\begin{figure*}[t]
\centering
    \centering

    \begin{minipage}{0.33\textwidth}
        \centering
        \includegraphics[width=\linewidth, trim= 0 0 0 0, clip]{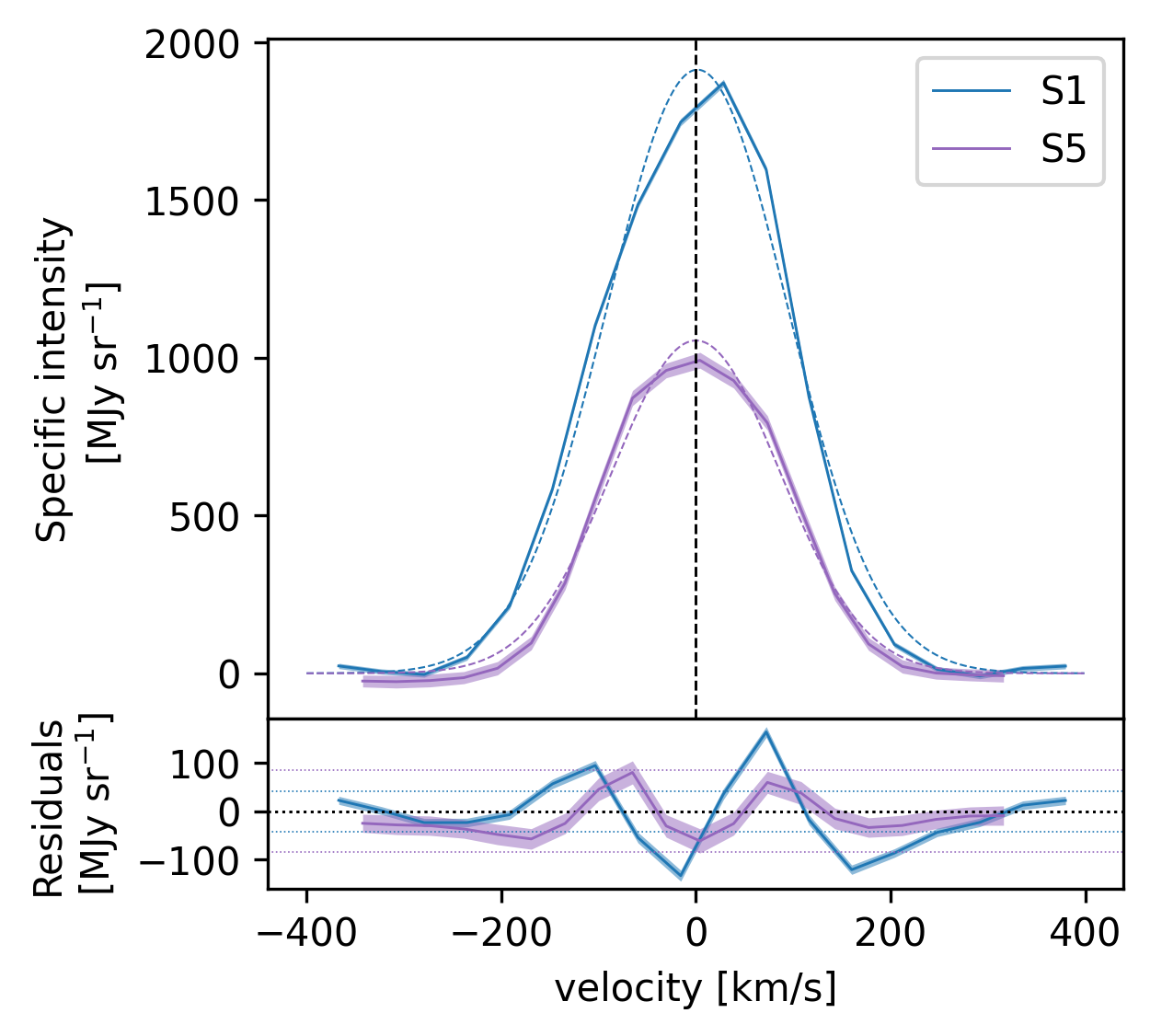}
        \put(-120, 130){\color{black}\text{CH1}} 
    \end{minipage}
    \hfill
    \begin{minipage}{0.33\textwidth}
        \centering
        \includegraphics[width=\linewidth, trim= 0 0 0 0, clip]{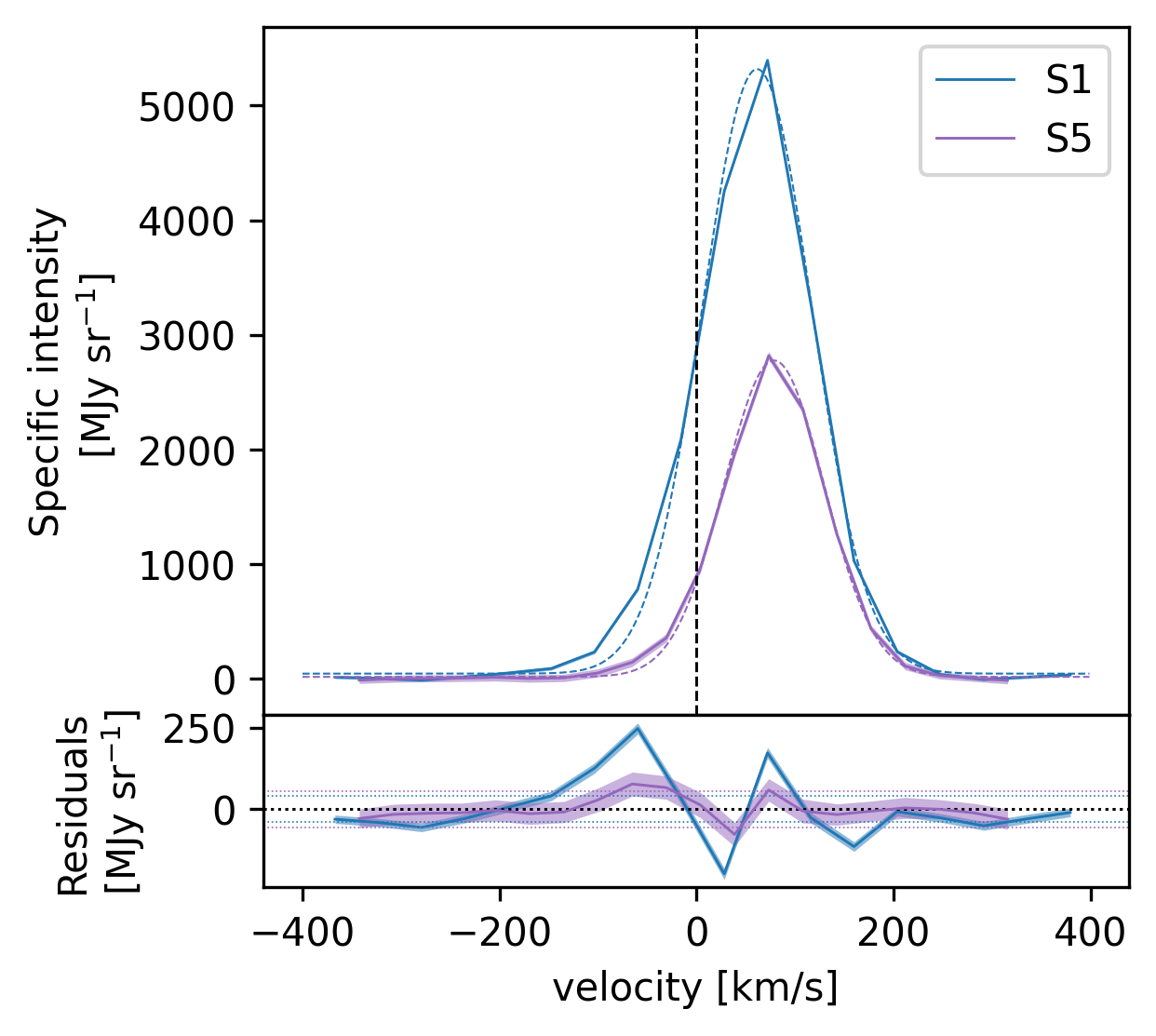}
        \put(-120, 130){\color{black}\text{North}} 
    \end{minipage}
    \hfill
    \begin{minipage}{0.33\textwidth}
        \centering
        \includegraphics[width=\linewidth, trim= 0 0 0 0, clip]{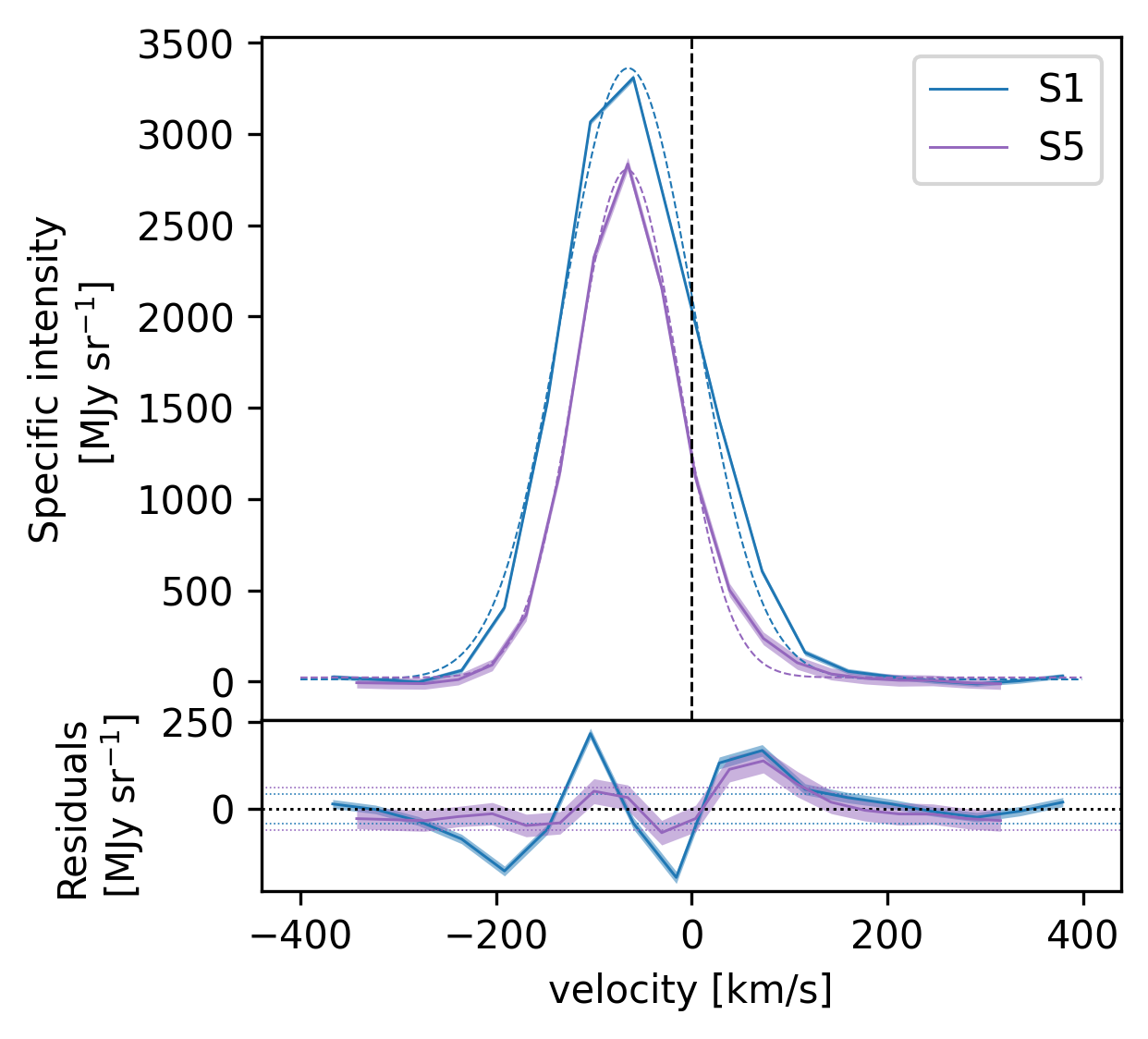}
        \put(-120, 130){\color{black}\text{South}} 
    \end{minipage}
    
    \caption{\Hmol\ 0--0 S(1) and S(5) line profiles averaged over the FoV of CH1 (left) and the two elliptical regions located at the base of the S-shaped distortion, as indicated in red in Fig.~\ref{fig:s5m2}, north (center) and south (right) of the AGN. Uncertainties are highlighted over the plots. The residuals shown beneath each profile trace the deviations from 1D-Gaussian fits with respect to a $3\sigma_\mathrm{STD}$ criterion (dotted horizontal line). Clear deviations are noticeable for both lines at $\pm100$~\kms, except for the S(5) on the FoV of CH1, where deviations are within the limits.}
    \label{fig:disp_profiles}
\end{figure*}

Although the rotational \Hmol\ line profiles averaged over the entire FoV are well reproduced by single Gaussian components, we note that, locally, the \Hmol\ lines exhibit slight asymmetries, particularly in the central regions. Figure~\ref{fig:disp_profiles} presents the spatially averaged line profiles of the 0--0 S(1) and 0--0 S(5) transitions extracted over the full FoV and over two elliptical regions located north and south of the AGN, along the line of nodes of the warped disk (regions highlighted in red in Fig.~\ref{fig:s5m2}) near the basis of the S-shaped distortion. The profiles show residuals from the single Gaussian fits exceeding the $\sim3\sigma_\mathrm{STD}$ level, with deviations centered around -100 and +100~\kms. This pattern may trace enhanced velocity dispersion resulting from the adiabatic compression of infalling gas onto the central regions of the CND \citep{vollmer_quenching_2013}. Alternatively, it could reflect the presence of a radially outflowing component \citep{alonso-herrero_miconic_2025}, possibly arising from inhomogeneous emission within an expanding shell or bubble, where projection effects lead to asymmetric contributions from the approaching and receding sides. Both types of velocity structures are commonly seen in numerical simulations of CNDs \citep{guo_2024} and in jet-driven outflows \citep[e.g.,][]{mukherjee_2018,mukherjee_2025} (see discussion in Sect.~\ref{subsec:H2ionized}).

\begin{table}[h]
   \caption{List of velocity dispersions for every \Hmol\ line, obtained via single Gaussian fit. }
\centering
\begin{center}

\begin{tabular}{c c c c c c}
\hline 
Line & \multicolumn{5}{c}{width (FWHM) [\kms]} \\
0--0 & Full & ND & ICND & (a) & (b) \\ \hline\hline
S(1) & 206 $\pm$ 5 & 124 $\pm$ 5 & 210 $\pm$ 5 &  78 $\pm$ 5 & 107 $\pm$ 5 \\
S(2) & 189 $\pm$ 5 & 127 $\pm$ 5 & 199 $\pm$ 5 &  65 $\pm$ 5 & 102 $\pm$ 5 \\ 
S(3) & 193 $\pm$ 3 & 161 $\pm$ 3 & 199 $\pm$ 3 &  82 $\pm$ 3 & 115 $\pm$ 3 \\ 
S(4) & 199 $\pm$ 4 & 169 $\pm$ 4 & 206 $\pm$ 4 &  83 $\pm$ 4 & 109 $\pm$ 4 \\ 
S(5) & 185 $\pm$ 3 & 163 $\pm$ 3 & 196 $\pm$ 3 &  92 $\pm$ 3 & 109 $\pm$ 3 \\ 
S(6) & 169 $\pm$ 3 & 165 $\pm$ 3 & 173 $\pm$ 3 &  65 $\pm$ 3 &  91 $\pm$ 3 \\ 
S(7) & 183 $\pm$ 3 & 169 $\pm$ 3 & 187 $\pm$ 3 &  62 $\pm$ 3 &  94 $\pm$ 3 \\ 
S(8) & 178 $\pm$ 3 & 172 $\pm$ 3 & 176 $\pm$ 3 &  59 $\pm$ 3 &  99 $\pm$ 3 \\ 
\hline
 \end{tabular}\\

 \end{center}
\tablefoot{Extractions are performed in the full FoVs (Full), the ND, the ICND, and respectively in the (a) and (b) hotspots (see Fig.~\ref{fig:moment 0 maps}).}
\label{table:lines_data}
\end{table}

\subsection{Excitation of the molecular gas}
\label{subsec:H2excitation}

We present our observational results on \Hmol\ excitation, comparing different regions, and on a spaxel-by-spaxel basis. The detailed physical modelling of the \Hmol\ excitation will be treated in the companion Paper II.

\subsubsection{Excitation diagrams}

\begin{figure}
\includegraphics[width=0.9\linewidth, trim=0 0 0 0, clip]{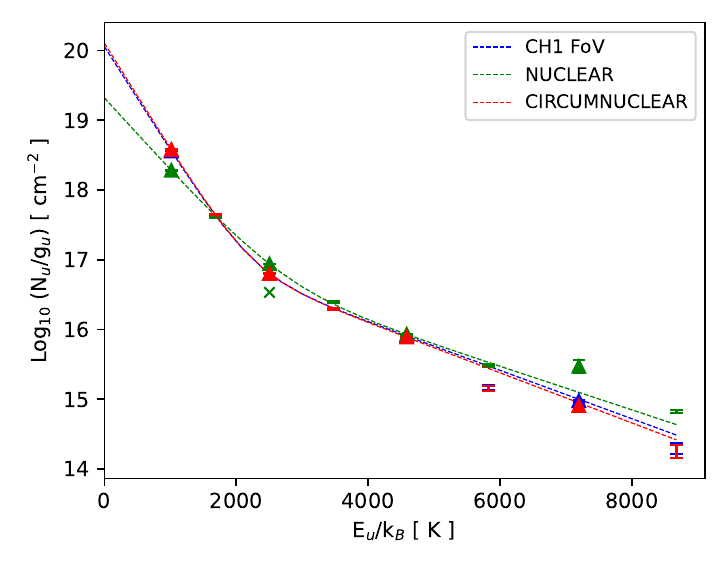}
\caption{Excitation diagrams for \Hmol\ extracted from the averaged cubes, over the full FoV of CH1 (blue), inside (green), and outside (red) a circle of radius $2\times$FWHM centered around the AGN. The triangles represent the column densities corrected for the OPR. The dashed lines represent the 2-temperature linear fits. 
The S(3) column density is corrected for extinction in the ND via the method presented by \citet{reefe_2025} (see App.~\ref{appendix:ext}). The uncorrected S(3) column density is indicated by the green cross.
The deviation of the S(7) from the fit in the ND is due to the difficult de-blending with the [MgVII] line at 5.51~\mum.
The results of the fits are listed in Table~\ref{tab:ed}.}

\label{fig:ed}
\end{figure}

We constructed \Hmol\ excitation diagrams from the line intensities averaged over three regions (tabulated in App.~\ref{appendix:spectra} Tab.~\ref{tab:flux}): the FoV of CH1, the ND, and the ICND. We assume that the ND emission is contained in a circle of radius $2\times$FWHM. The excitation diagrams, which  are shown in Figure~\ref{fig:ed}, display the logarithm of the upper level column densities divided by their statistical weights (degeneracy) $\log N_u / g_u$ as a function of the energy of the upper level of the transition in kelvin, $E_u/k_B$. The diagrams show a typical curvature, indicative of a distribution of gas temperatures. 

We find that the diagrams in the three different regions can be fitted by a simple two-temperature model, and we use the PDRTPY routine to derive temperatures, column densities, and \Hmol\ ortho-to-para ratios \citep{pound_photodissociation_2022}. At molecular cloud densities, typically $n_{\rm H} = 10^3$~\cc, the first levels ($J_u=1-5$) of \Hmol\ are thermalized, so they can be used to estimate the total column density $N_{\rm H_2}^{\rm (lin)}$ and the warm gas temperature component, which we find to be within a range of $\sim 250- 420$~K. The higher levels $J_u > 5$, corresponding here to transitions from 0--0~S(4) to S(8), are non-thermalized, and we find excitation temperatures ranging from $\sim 900 - 1400$~K. The routine also corrects column densities of the odd J transitions by fitting the \Hmol\ ortho-to-para ratio (OPR). 
The column density of the $J_u = 3$ level (corresponding to the S(1) line) over the full FoV is between $(2.7-7.6)\times10^{19}~\mathrm{cm^{-2}}$, so it is well below the critical column density ($3.2\times10^{25}~\mathrm{cm^{-2}}$ for $J_u = 3$). The \Hmol\ lines are thus optically thin in the ICND.

Estimating the optical depth of the ICND is difficult because the strong mid-IR continuum of the unresolved nucleus outshines part of the ICND, and the 9.7~\mum\ silicate absorption feature is faint. Fitting the \textit{Spitzer}/IRS spectrum with the PAHFIT tool \citep{smith_2012}, which assumes a mixing of the dust and the emitting gas, \citet{Ogle_2010} estimate an optical depth of 0.5 to 0.8 in the 5-30~\mum\ range, except in the silicate absorption band, near the S(3) line, where it reaches a high value of 2.2. This is an upper limit for the ICND because, as shown in Fig.~\ref{fig:spectrum}, the silicate absorption feature is located in the ND, and is much less prominent outside. 
We hence estimate the extinction of the S(3) line in the ND from the S(4)/S(3) line ratios, following the \citet{reefe_2025}, which we explain in Appendix~\ref{appendix:ext}. We find an optical depth $\tau_\mathrm{9.7\mu m}=0.9$, which is in agreement with the estimates based on the modeling of the dust continuum \citep{alexander_99, karovska_2003}. The S(3) line in the nuclear excitation diagram in Figure~\ref{fig:ed} (green) is corrected accordingly.

The results of the fits are listed in Table~\ref{tab:ed}, along with the \Hmol\ and total gas masses integrated over the respective FoVs for the different regions. These results show heterogeneous excitation among the different observation regions, with excitation temperatures peaking in the ND at $\sim 420$~K for the thermalized levels and to almost 1400~K for the higher levels. The OPR is estimated around 2 in the ICND and around 3 in the ND. 

The fit yields a total column density for \Hmol\ at 300~K around $8.2\times10^{20}~\mathrm{cm^{-2}}$ which drops in the ND by a factor 4. This is consistent with the central cavity observed in the S(1) map, since this line traces most of the warm molecular mass.

\begin{table}[!t]
\caption{Physical parameters fitted from the excitation diagrams}
    \centering
    \begin{tabular}{c c c c}
    \hline
                &   CH1    &  ND   & ICND     \\
        \hline
        \hline

FoV [arcsec$^2$] & 28.5 & 6.3 & 22.2 \\
$T_{\mathrm{hot}}$ [K] & $ 1252 \pm 45 $ & $ 1391 \pm 139 $ & $ 1209 \pm 44 $ \\

$T_{\mathrm{warm}}$ [K] & $ 291 \pm 7 $ & $ 416 \pm 12 $ & $ 286 \pm 7 $ \\

OPR & $ 2.0 \pm 0.1 $ & $ 2.9 \pm 0.1 $ & $ 2.0 \pm 0.1 $ \\

$N_{\mathrm{H_2}}^{\mathrm{(lin)}}$ [$10^{20}$ cm$^{-2}$] & $ 8.2 \pm 0.9 $ & $ 2.2 \pm 0.2 $ & $ 9.0 \pm 1.0 $ \\
$M_{\mathrm{H}_2}^{\mathrm{(lin)}}$ [$10^{5}$ M$_\odot$] & $ 1.4 \pm 0.2 $ & $ 0.08 \pm 0.01 $ & $ 1.2 \pm 0.1 $ \\

$M_{\mathrm{H}_2}^{\mathrm{(pow)}}$ [$10^{5}$ M$_\odot$] & $ 29 \pm 11 $ & $ 0.46 \pm 0.09$ & $ 25 \pm 9 $ \\

$M_{\mathrm{gas}}^{\mathrm{(pow)}}$ [$10^{5}$ M$_\odot$] & $ 39 \pm 15 $ & $ 0.63 \pm 0.12 $ & $ 35 \pm 13 $ \\

$M_{\mathrm{H_2}}^{\mathrm{(100K)}}$ [$10^{5}$ M$_\odot$] & $ 5.6 \pm 1.4 $ & $ 0.56 \pm 0.11 $ & $ 5.0 \pm 1.2 $ \\

$M_{\mathrm{gas}}^{\mathrm{(100K)}}$ [$10^{5}$ M$_\odot$] & $ 7.6 \pm 1.9  $ & $ 0.76 \pm 0.15 $  & $ 6.8 \pm 1.6 $ \\

    \hline
    \end{tabular}
     \tablefoot{ The fit are performed over the full FoV of CH1, inside (ND), and outside (ICND) a circle of radius $2\times$FWHM centered around the AGN. The fits for the excitation temperatures $T_\mathrm{hot}$ and $T_\mathrm{warm}$ were performed with a two-linear-component method (Figure~\ref{fig:ed}) and correcting for the ortho-to-para ratio (OPR) between odd and even transitions.
    The \Hmol\ column densities $N_\mathrm{H_2}^\mathrm{(lin)}$ are integrated over the different regions to obtain the respective masses $M_\mathrm{H_2}^\mathrm{(lin)}$. $M_\mathrm{H_2}^\mathrm{(pow)}$ is the mass estimated via power-law fit of the excitation diagrams (see Sect.~\ref{sec:mass}). $M_\mathrm{H_2}^\mathrm{(100K)}$ is the mass estimated with the same method but setting a limit on the minimum temperature of the fit at 100~K (see App.~\ref{appendix:h2p}). $M_\mathrm{gas}^\mathrm{(pow)}$ and $M_\mathrm{gas}^\mathrm{(100K)}$ are the masses corrected by a factor 1.36 to account for He.}
    \label{tab:ed}
\end{table}

\begin{figure*}[h]
    \centering
    \includegraphics[width=0.7\linewidth,trim= 0 10 0 0,clip]{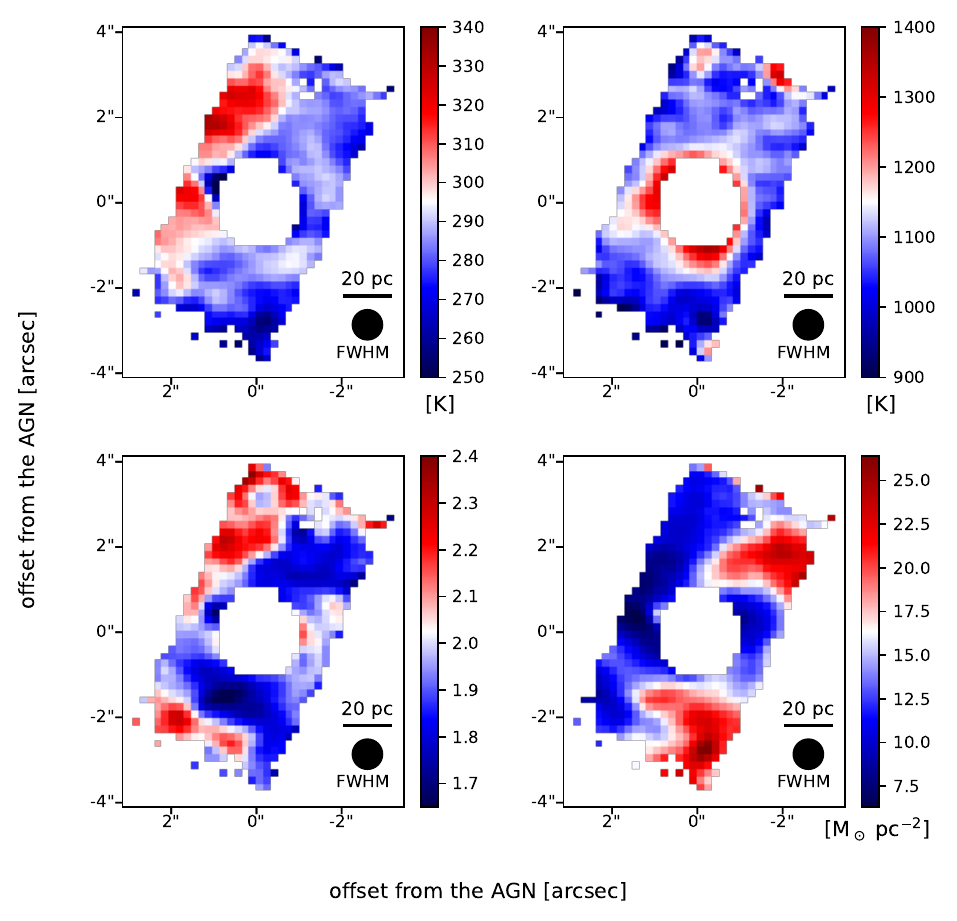}
    \put(-310, 320){\color{black}\text{T$_\mathrm{warm}$}}
    \put(-140, 320){\color{black}\text{T$_\mathrm{hot}$}}
    \put(-310, 155){\color{black}\text{OPR}}
    \put(-147, 162){\color{black}\text{surf. mass}}
    \caption{Maps of physical parameters constructed with the \href{https://github.com/mpound/pdrtpy}{PDRTPY} routine \citep{pound_photodissociation_2022} via spaxel-by-spaxel two-linear-component fit of the excitation diagrams: $T_\mathrm{warm}$ map of the warm temperature fit component (upper right); $T_\mathrm{hot}$ map of the hot temperature fit component (upper left); \Hmol\ ortho-to-para ratio (OPR) map (bottom left); surface mass map (bottom right). The maps are convolved to the resolution of the \Hmol\ 0--0 S(1) map and reprojected to the spaxel grid of CH2. The FoV is limited to the spaxel coverage of the smallest map. The \Hmol\ 0--0~S(6) and S(8) maps were excluded due to the high number of flagged spaxels. }
    \label{fig:ex_maps}
\end{figure*}

\subsubsection{Spatial variations of \Hmol\ temperature and ortho-to-para ratios}

We present here the warm and hot excitation temperature maps (in K), as well as the OPR map, generated using the PDRTPY routine \citep{pound_photodissociation_2022} on the convolved and reprojected maps, as shown in Figure~\ref{fig:ex_maps}.
The two upper panels display the excitation temperature maps. The hot regime map (right) reveals a hot central region surrounding the AGN with excitation temperatures peaking at 1400~K and decreasing outward into the ICND. This morphology identifies the ND where the excitation of \Hmol\ is likely driven by UV/X-ray radiation from the AGN \citep{borkar_multiphase_2021,  vollmer_circumnuclear_2022}, with most of the emission arising from higher rotational lines, 0--0 S(4) to S(8), as well as ro-vibrational line emission \citep{neumayer_central_2007}. In contrast, the warm regime map (left) shows a colder nuclear cavity, with temperatures in the ICND up to 300~K. The warm map also reveals a warmer region on the eastern (left) side of the disk, with temperatures up to $\sim$50~K higher than the western side. This  region overlaps with the area identified by \citet{alonso-herrero_miconic_2025} as showing enhanced [Ne III]/[Ne II] ratios and by \citet{munoz_2025} as exhibiting correlated mid-IR spectral properties, all pointing to distinct physical processes operating there.

The OPR map (bottom left) exhibits pronounced spatial variations across the FoV, with values ranging from 1.6 to 2.4, and never reaching the LTE value of 3. The OPR attains its highest values, exceeding 2.2, in the eastern and north-eastern regions, coinciding with the warmer area identified in the temperature map (top left). This side of the map corresponds to the portion of the disk where the jet is oriented toward the observer. In contrast, the ratio decreases below 1.9 in two distinct regions located north and south of the AGN. Notably, the northern low-OPR region overlaps with the (a) hotspot (see Fig.~\ref{fig:moment 0 maps}).

\subsection{Continuum emission and \Hmol-to-continuum ratio}
\label{subsec:continuum}

\begin{figure}[h!]
    \centering

    \includegraphics[width=1\linewidth]{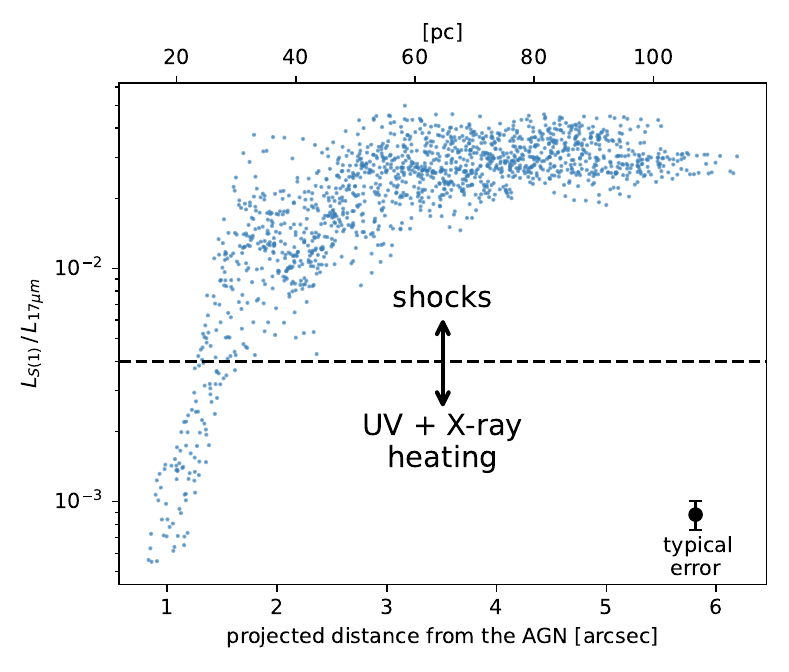}
 
   \caption{Spaxel-by-spaxel plot of the ratio between the \Hmol\ 0--0 S(1) luminosity and the monochromatic continuum luminosity $\nu L_\nu$ at 17~\mum\ (with $L_\nu$ spectral luminosity). The ratio increases with the projected distance from the AGN up to 30~pc, pointing at a stronger non-radiative excitation of  \Hmol\ in the outer parts of the ICND. The dashed horizontal line indicates the UV PDR limit from \citet{Guillard_2012} rescaled using the $L_\mathrm{PAH_{7.7\mu m}}/L_\mathrm{17\mu m}$ ratio of the PAH$_{7.7\mu m}$ and the continuum at 17~\mum\ (averaged over the ICND).}
    \label{fig:17}
\end{figure}

\begin{table*}[h]
\hspace{-0.5cm}
\caption{Luminosity ratios between the sum of the mid-IR \Hmol\ lines (S(1) to S(8)) and the monochromatic continuum at 24~\mum, 17~\mum, the PAH$_{7.7\mu\mathrm{m}}$ feature, and the X-ray luminosity (2-10 keV).}
    \centering
        \begin{tabular}{c c c c c c c}
    \hline
        Ratio & Full & ND  & ICND & CH1 & (a) & (b) \\ 
         
        \hline
        \hline

Log$_{10}\left(L_{\Hmol}/L_{24\mu\mathrm{m}}\right)$ 
        & -2.60 $\pm$ 0.04 & -3.73 $\pm$ 0.04 & -1.57 $\pm$ 0.04 & 
        -2.66 $\pm$ 0.04  & -1.78 $\pm$ 0.04 & -2.02 $\pm$ 0.04 \\

Log$_{10}\left(L_{\Hmol}/L_{17\mu\mathrm{m}}\right)$ & 
        -2.43 $\pm$ 0.04 & -3.57 $\pm$ 0.04 & -1.38 $\pm$ 0.04 &
        -2.52 $\pm$ 0.04 & -1.46 $\pm$ 0.04 & -1.73 $\pm$ 0.04 \\

Log$_{10}\left(L_{\Hmol}/L_\mathrm{PAH_{7.7\mu \mathrm{m}}}\right)$ 
        & -0.48$\pm$ 0.03 & -1.05$\pm$ 0.02 & -0.38$\pm$ 0.03 & -- & -- & --  \\

Log$_{10}\left(L_{\Hmol}/L_\mathrm{X(2-10~keV)}\right)$ 
        & -2.36 $\pm$ 0.04 &  -3.29 $\pm$ 0.04 & -1.89 $\pm$ 0.04 & -- & -- & -- \\

\hline

    \end{tabular}
    \tablefoot{ Extractions are from the entire FoV (Full), the ND located within an aperture of radius $2\times$FWHM, the ICND outside, the area covered by the FoV of CH1, as well as the (a) and (b) hotspots (see Section~\ref{sect:morph} and Figure~\ref{fig:moment 0 maps}). Each \Hmol\ line luminosity is extracted in its respective spectral MRS channel. The monochromatic continuum luminosities are extracted from the sub-cubes (see Sect.~\ref{subsec:continuum}).  
    The PAH$_{7.7\mu\mathrm{m}}$ fluxes are provided by \citet{pantoni_2026}. The $L_\mathrm{X(2-10~keV)}$ values were obtained from \citet{Ogle_2010}. The spatial resolution for the PAH$_{7.7\mu\mathrm{m}}$ and X-ray observations is not sufficient to perform smaller aperture extractions for the CH1, (a) and (b) regions} 
    \label{tab:ratios}
\end{table*}

Table~\ref{tab:ratios} provides the luminosity ratio between the mid-IR \Hmol\ lines and the monochromatic continuum at 24~\mum, to allow a comparison with previous studies \citep{Ogle_2010, Guillard_2012, u_goals-jwst_2022}. We compute the continuum luminosity $L_{24\mu\mathrm{m}}$ as the specific intensity of the cube at 24~\mum, integrated over the entire FoV, multiplied by the frequency $c/24$~\mum, where $c$ is the speed of light. The ratio $L_\mathrm{H_2}$/$L_{24\mu\mathrm{m}}$ presents a difference of almost 2 orders of magnitude between the ICND and the ND, where the continuum is stronger (see Table~\ref{tab:ratios}). 

Figure~\ref{fig:17} presents the spaxel-by-spaxel $L_{S(1)}/L_{17\mu\mathrm{m}}$ luminosity ratio between the \Hmol\ 0–0 S(1) line and the adjacent monochromatic continuum at 17~\mum. We find that this ratio has a strong dependence with the projected distance from the AGN in the innermost part of the CND (<1.8" or 30~pc), and then is constant with radius. The radial variation of the $L_{S(1)}/L_{17\mu\mathrm{m}}$ ratio is primarily governed by the continuum intensity, as the S(1) surface brightness varies by less than a factor of two over this distance range. UV PDR models predict a maximum value for the \Hmol-to-PAH$_\mathrm{7.7\mu m}$ luminosity ratio of 0.04 (including X-ray pumping) \citep{Guillard_2012}. Given the ratio $L_\mathrm{PAH_{7.7\mu m}} / L_\mathrm{17\mu m}=0.1$ between the PAH$_\mathrm{17\mu m}$ and the continuum at 17~\mum\ averaged over the ICND, we rescale the PDR limit to $4\times10^{-3}$. The total \Hmol-to-17~\mum\ continuum ratio exceeds this limit for all spaxels >1.3" (or 24~pc), therefore we rule out UV+X-ray heating as the dominant excitation mechanism of the \Hmol\ emission in the ICND (see discussion in Sect.~\ref{subsec:H2excit_mechanism} and \ref{subsec:H2excit_turbulence}).

We also provide continuum maps in Appendix~\ref{appendix:maps} (Figure~\ref{fig:continuum}). Table~\ref{table:continuum} indicates the spatial FWHM of the centroid of the continuum emission estimated via 2D-Gaussian fit of the maps. The measured widths are compatible with the expected MRS spatial resolution \citep{law_3d_2023}, suggesting that the emitting source is unresolved. 

\begin{figure*}[h!]
    \centering

    \begin{overpic}[width=0.95\linewidth]{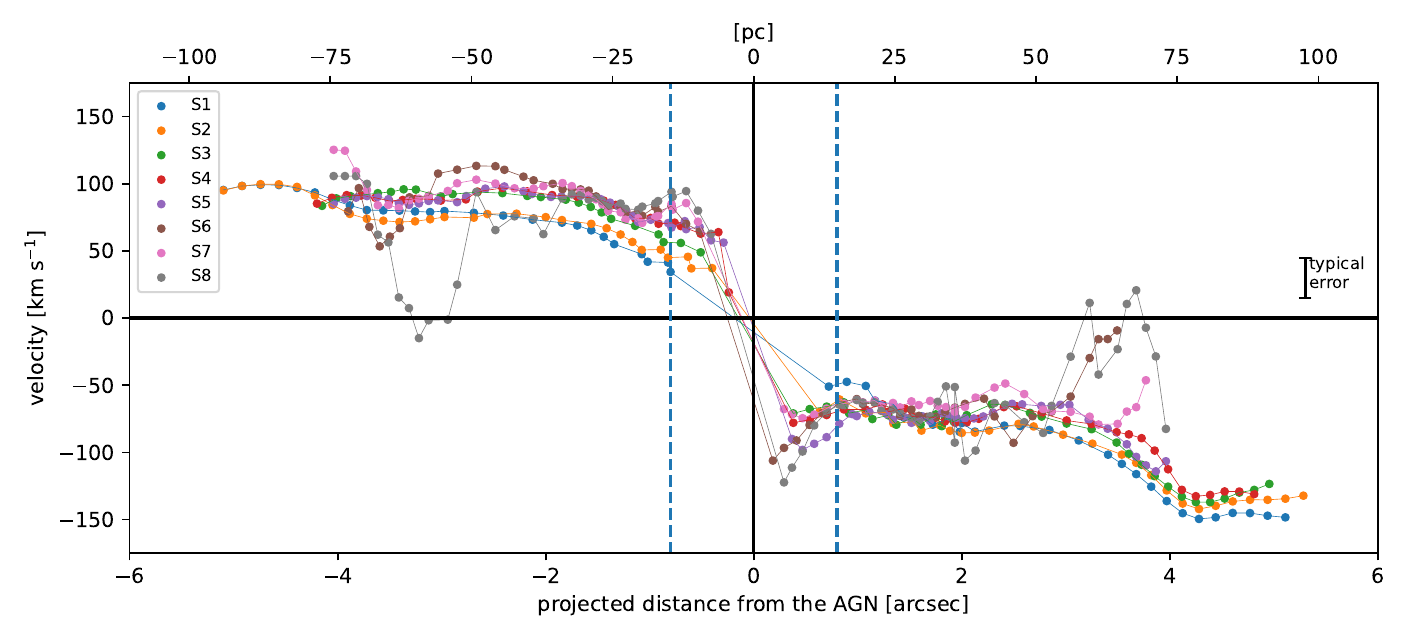}

        \put(10, 7){
            \includegraphics[width=0.13\linewidth, trim=115 35 95 40, clip]{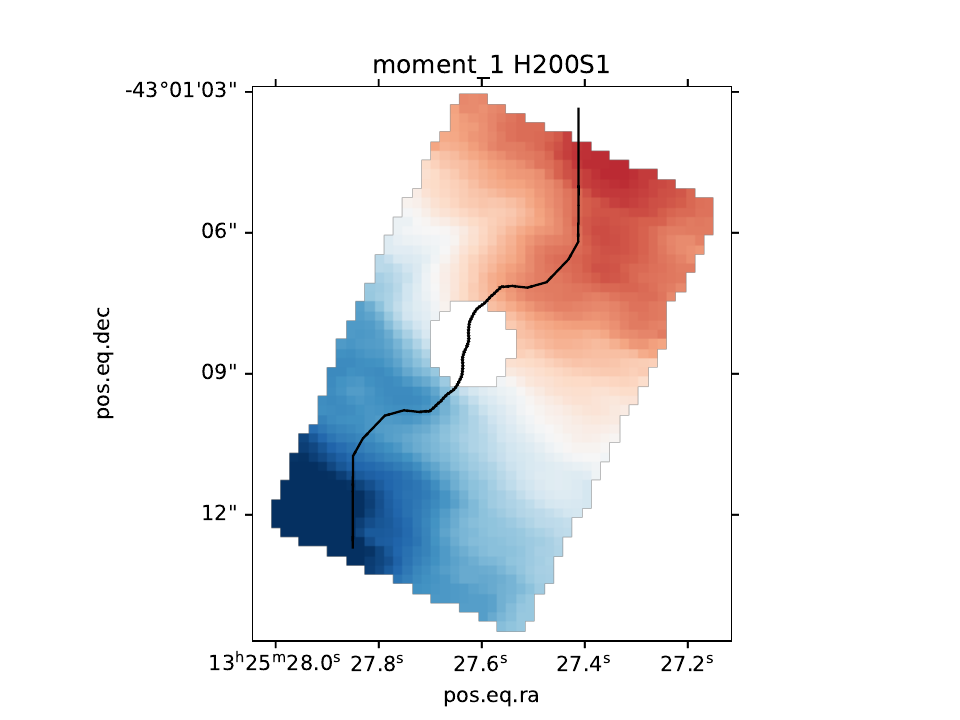}
        }
        \put(22, 25){\color{black}\vector(-1,0){5}}   
        \put(17, 26){\color{black}\text{NORTH}}   
        
    \end{overpic}
    \caption{P-V diagrams of the \Hmol\ gas extracted from the rotational line velocity maps. The path chosen follows the line of the nodes of the warped disk model from \citet{neumayer_central_2007} to maximize the projected velocity component. The small quadrant shows the S(1) velocity map with the path of extraction for reference. Every point in the diagram is averaged on a square of nine spaxels (120~pc$^2$) following the black line. A typical 1$\sigma$ error bar of 30~\kms\ is given on the right. The blue dashed lines represent the expected radius of influence of the SMBH given the mass estimated by \citet{neumayer_central_2007}.}
    \label{fig:pv}
\end{figure*}

\subsection{Molecular mass gas and dynamical mass}\label{sec:mass}

We present here our estimate of the total \Hmol\ mass. The bottom right panel of Figure~\ref{fig:ex_maps} shows the surface density map, in \Msun\,pc$^{-2}$, generated using the \texttt{\href{https://github.com/mpound/pdrtpy}{PDRTPY}} routine \citep{pound_photodissociation_2022}. Since the column density is dominated by the lowest energy levels, the map closely resembles the morphology of the 0-0~S(1) line, with the hotspots (a) and (b) clearly visible, and bearing respectively 30\% and 20\% of the whole mass, while the eastern region shows a column density almost one order of magnitude lower than the hotspots. From this map, we integrate a total molecular hydrogen mass of $M_\mathrm{H_2}^\mathrm{(map)} = 8.1 \times 10^4$~\Msun\, over an FoV of 19.8~arcsec$^2$. This result is consistent with the mass $M_\mathrm{H_2}^\mathrm{(lin)} = 1.4 \times 10^5$~\Msun\, obtained from the linear fit of the excitation diagrams reported in Table~\ref{tab:ed} for the FoV of CH1, once we account for the fact that the FoV of the surface density map is $~40\%$ smaller due to flagged spaxels. A comparison with the cold mass gas estimate from \citet{espada_disentangling_2017} is provided in Section~\ref{sec:comp_mass}.

Since the S(0) line is outside the MRS spectral coverage, the mass estimate from the two-linear-component fit of the excitation diagram is limited to the $\sim 300$~K warm gas. We used the \href{https://github.com/astrolojo/H2Powerlaw}{H2Powerlaw}\footnote{https://github.com/astrolojo/H2Powerlaw} tool from \citet{togi_2016} to perform a power-law fit of the excitation diagrams to extrapolate the mass of  \Hmol\ to lower temperatures. The power-law fits are presented in Appendix~\ref{appendix:h2p} Figure~\ref{fig:ex_pl}. These fits estimate the mass $M_\mathrm{H_2}^\mathrm{(pow)}$ for \Hmol\ to the lower fitting temperatures $T_\mathrm{low}$ of 110~K in the nucleus and 60~K in the ICND. Since \Hmol\ does not radiate in IR at 60~K, we increase the lower limit of the fit $T_\mathrm{low}$ to 100~K, and hence find a mass $M_\mathrm{H_2}^\mathrm{(100K)}$ about 5 times lower than $M_\mathrm{H_2}^\mathrm{(pow)}$ in the ICND. 

These results are listed in Table~\ref{tab:ed}.     

Figure~\ref{fig:pv} shows the position–velocity (PV) diagram extracted along the line of the nodes defined by the warped disk model from \citet{neumayer_central_2007}. We follow the line of the nodes to ideally maximize the projected line-of-sight bulk velocity component. The wavelength solution calibration of \citet{argyriou_jwst_2023} through the MRS wavelength range was used to compute the uncertainties on velocities. Due to masking of the central spaxels in our maps, we are unable to trace the gas dynamics within the black hole’s sphere of influence, which is marked by the vertical blue lines at 0.8" (15~pc) from the center \citep{neumayer_central_2007}. 

Assuming a Newtonian approximation, the dynamical mass enclosed within a sphere of radius $r$ is given by: 
\begin{equation}
M_\mathrm{dyn} = \frac{r}{G} \left[\frac{V}{ \cos(i)}\right]^2\
\end{equation}
where $G$ is the gravitational constant, $V$ is the projected line-of-sight velocity, and $i$ is the inclination angle. We adopt a radius of $r \sim 4''$ and estimate $V$ as the average S(1) velocity in the radial interval between $3.5''$ and $4.5''$. Position angle values are taken from \citet{neumayer_central_2007} and span around a median PA$_\mathrm{nodes}\sim155^\circ$. This yields a dynamical mass of $ (5\pm0.4)\times10^8$~\Msun. Such estimate encompasses the contributions from \Hmol, stellar mass, and the central SMBH.

\section{Discussion}

\subsection{High \Hmol-to-dust IR emission ratio: evidence for shock excitation?}
\label{subsec:H2excit_mechanism}

We discuss here how the results that we presented can be interpreted as indicators of different sources of excitation. 
\citet{Ogle_2007, Ogle_2010} define an empirical threshold of $L_\mathrm{H_2}/L_\mathrm{24\mu m}=0.02$ between \Hmol\ line emission and continuum at 24\mum\ to classify galaxies that present significant non-radiative heating of the gas (so-called MOHEGs). The results presented in Section~\ref{subsec:continuum} show that this limit is exceeded in all regions studied in this work, except in the nuclear region. 

\citet{Ogle_2010} revisit this criterion, as it does not account for the variation in AGN contribution to the continuum emission and propose the ratio $L_\mathrm{H_2}$/$L_{\mathrm{PAH_{7.7\mu\mathrm{m}}}}$ as a much better tracer of the relative contribution of mechanical heating with respect to radiative heating.
Table~\ref{tab:ratios} presents the ratios computed from the PAH$_{7.7\mu\mathrm{m}}$ flux provided by \citet{pantoni_2026} using PAHDecomp \citep{donnan_2023, donnan_2024}. The ratio is above the limit of 0.04 expected from PDR models \citep{Guillard_2012} in all regions. Although the PAH feature in the ND does not emerge clearly from the continuum, we estimate a $L_\mathrm{H_2}$/$L_{\mathrm{PAH_{7.7\mu\mathrm{m}}}}$ ratio in the ICND that is 4 times higher compared to the ND. This is consistent with \citet{israel_outflow_2017} who find shock-excited gas at <1000~K from the 1--0 S(J) line ratios observed by \textit{Herschel}. Such ratios strongly suggest that collisional excitation from shocks in the ICND provides a major source of molecular hydrogen excitation. 

\citet{espada_disentangling_2017} support the same conclusion for CO, indicating the non-axisymmetric  morpho-kinematics and the resulting torque exerted on the gas as possible causes of loss of angular momentum. The same argument can be extended to the non-axisymmetric \Hmol\ morpho-kinematical features, reinforcing the importance of shocks in energy and angular momentum dissipation for the molecular gas.

Strong H$_2$-to-PAH ratios have been observed in 3C~293 by \citet{riffel_bah2_2025}, in IC5063 by \citet{dasyra_2024}, and will be discussed for other radio loud AGNs by Riffel et al in prep, particularly in regions where the velocity dispersion is high (up to $\sim500$~\kms).

\subsection{\Hmol\ emission from dissipation of turbulent mechanical energy}
\label{subsec:H2excit_turbulence}

We first examine the role of turbulent heating in maintaining the observed \Hmol\ emission. 
Following \citet{guillard_sq_2012}, the turbulent heating rate required to maintain the \Hmol\ luminosity can be estimated from the energy dissipation rate, expressed as the luminosity-to-mass ratio \( L_\mathrm{H_2}/M_\mathrm{H_2} \), where $M_\mathrm{H_2}$ is the mass of \Hmol\ at 300~K. 
If the emission is entirely powered by turbulent dissipation, the following relation holds:
\begin{equation}
    \frac{3}{2}\frac{\sigma_T^3}{l} = \frac{L_\mathrm{H_2}}{M_\mathrm{H_2}},
\end{equation}
where the left-hand side of the equation is the turbulent heating rate \citep{maclow_1999}, with \( \sigma_T \) the turbulent velocity dispersion, and \( l \) is the characteristic size of the region over which the dispersion is measured.

\citet{Ogle_2010} report dissipation rates between 1 and 33~\(\mathrm{L_\odot}/\mathrm{M_\odot}\) for their sample of MOHEGs. 
Over the CH1 FoV, we derive for Centaurus~A a dissipation rate of \( L_\mathrm{H_2}/M_\mathrm{H_2}^\mathrm{(lin)} = 4.43~\mathrm{L_\odot}/\mathrm{M_\odot} \), therefore within the MOHEG range. 
Adopting $ l_\mathrm{FoV} = 4.5'' $, corresponding to the width of the FoV in CH1, we infer a turbulent velocity dispersion (FWHM) of 270~\kms\ required to sustain the observed \Hmol\ emission. This is $~40\%$ higher than the observed mean velocity dispersion (see Table~\ref{table:lines_data}). However, at the scale of the FoV, the complex gas kinematics and disk geometry make it difficult to isolate quantitatively the turbulence from the rotational and radial bulk flow components in the velocity dispersion field. 

Locally, at the PSF scale $l_\mathrm{PSF}$, we can identify regions where the rotational component is small, especially close to the minor kinematic axis along the \Hmol\ filaments. Here the turbulent motions dominate the velocity field. Assuming $ l_\mathrm{PSF} = 0.3'' $ (6~pc), corresponding to the PSF size in CH1, we compute the turbulent heating rate for velocity dispersions between 70 and 90~\kms, as observed along the coherent spiral streamer in the S(5) dispersion map (Fig.~\ref{fig:s5m2}). 
This yields turbulent heating rates of 1.7 and 2.6~\(\mathrm{L_\odot}/\mathrm{M_\odot}\), respectively.
Over the same \( l_\mathrm{PSF} = 0.3'' \) aperture along the spiral streamer, we measure a luminosity-to-mass ratio of \( L_\mathrm{H_2}/M_\mathrm{H_2}^{(l_\mathrm{PSF})} \sim 2.7~\mathrm{L_\odot}/\mathrm{M_\odot} \). 
At the 10~pc scale, the dissipation rate thus matches the estimated turbulent heating rate, suggesting that the dissipation of mechanical turbulent energy significantly contributes to the excitation of the molecular gas. This is in agreement with theoretical models and observations of the CND in the galactic center of NGC~1068 \citep{vollmer_circumnuclear_2022}.

\subsection{X-ray heating and cosmic ray heating of the \Hmol\ gas}

We discuss here the impact of X-ray photons from the AGN corona and of comic rays (CR) on \Hmol\ excitation. Using  X-ray-dominated region (XDR) models \citep{maloney_1996}, \citet{Ogle_2010} estimated a maximum \Hmol-to-X-ray luminosity ratio of $L_\mathrm{H_2}/L_\mathrm{X(2-10~keV)}\lesssim 0.01$ in X-ray dominated environments. This value assumes that all the X-ray flux is absorbed by the XDR, and that the fraction of the absorbed X-ray flux that goes into gas heating by photoelectrons is high (40\%), but it could be a factor of 2 lower. The Chandra X-ray luminosity of the central, unresolved, point-source is $L_\mathrm{X(2-10~keV)}=5.01\times10^{41}~\mathrm{erg\,s^{-1}}$ \citep{Ogle_2010} Given Chandra's PSF profile \citep{kraft_2002}, less than 30\% of $L_\mathrm{X(2-10~keV)}$ can come from the ICND (within 1.3" or 24~pc from the AGN). This yields a $L_\mathrm{H_2}/L_\mathrm{X(2-10~keV)}$ ratio of 0.001 (lower limit) in the ND and 0.013 (upper limit) in the ICND. The former ratio is below the upper threshold of the XDR models, while the latter exceeds it, which suggests that radiative heating dominates in the ND, while additional heating mechanisms are required in the ICND, consistently with the results shown in Figure~\ref{fig:17}.  A detailed modeling of the contribution of XDRs to the \Hmol\ line emission is out of the scope of this paper. 

For CR heating of the \Hmol\ gas, we can estimate the ionization rate $\zeta$ required to sustain the \Hmol\ line emission. Assuming an ionization energy of 4~eV per ionized molecule \citep[MEUDON code,][]{lepetit_2006}, given the mass of an \Hmol\ molecule $m_\mathrm{H_2}$, and based on the luminosity-mass ratio, we find $\zeta=m_\mathrm{H_2}\ (L_\mathrm{H_2}/M_\mathrm{H_2}^\mathrm{(lin)})\ /\ 4\mathrm{eV}\sim4.5\times10^{-12}~\mathrm{s^{-1}}$. This result is 5 orders of magnitude higher than the typical ionization rate in the Milky Way \citep{shaw_2006}, 1 order of magnitude above the average found by \citet{Ogle_2010} for MOHEGs, and more than 2 orders of magnitude above the ionization rate observed in the central molecular zone of the Galactic center \citep{ravikularaman_cosmic_2025}. CR heating is thus unlikely to be the main powering source of the \Hmol\ line emission over the scale of Cen~A's ICND.

\subsection{Comparison of total molecular gas mass with Spitzer and ALMA}\label{sec:comp_mass}

We compare our gas mass estimates with those from previous studies.
Using Spitzer observations integrated within a 13.7 arcsec$^2$ field of view, \citet{Ogle_2010} derived a total \Hmol\ column density of $5\times10^{21}~\mathrm{cm^{-2}}$ by fitting the excitation diagram with three temperature components based on transitions from S(0) to S(7). Their result, rescaled to the MIRI FoV (CH1), assuming a homogeneous distribution, yields a total mass of $1.1\times10^{6}~\mathrm{M_\odot}$ for the  $\sim150$~K component in the excitation diagram , which is consistent with our estimate  of $M_\mathrm{gas}^\mathrm{(100K)}\sim0.8\times10^6$~\Msun.  
From CO observations, \citet{espada_disentangling_2017} reported a total gas mass of $9\times10^{7}~\mathrm{M_\odot}$ over a much larger field of view of 144 arcsec$^2$.  We extract the flux of the C0(3-2) line, integrating over their ALMA map, restricted to the region covered by the MIRI FoV (CH1). Using the same conversion factor of 0.1 they use between the CO(1-0) and the CO(3-2) lines and a factor $X_{\rm CO}=4\times10^{20}$~cm$^{-2}$~K~\kms \ \citep[for the nuclear region of Cen~A][]{israel_molecular_2014,miura_2021}, we find a total gas mass of 1.7$\times10^6$~\Msun. This result is about 2 times larger compared to our $M_\mathrm{gas}^\mathrm{(100K)}$ estimate. This discrepancy may be due to an overestimated $X_{\rm CO}$ factor or to the fact that ALMA observations probe gas that is colder than 100~K, which we are not detecting with MRS.

\subsection{Comparison between \Hmol\ and the ionized gas}\label{subsec:H2ionized}

Several results highlight morphological and kinematical differences between \Hmol\ and the ionized gas. The ionized line emissions peak in a cone-like structure aligned with the jet \citep{alonso-herrero_miconic_2025}, whereas \Hmol\ emission is more diffuse across the ICND. As shown in Figure~\ref{fig:moment 0 maps} (bottom left panel), we find similar contrasts between the [Ne{\sc \ vi}] and \Hmol\ morphologies: the [Ne{\sc \ vi}] emission is concentrated around the AGN, and extended along the jet axis, with no trace of the low-dispersion filamentary structures observed in \Hmol. The elongated morphology of the [O {\sc iv}] and [Ne {\sc v}] lines perpendicularly to the ICND and indicative of an outflow was also noted by \citet{quillen_2008}, from \textit{Spitzer}/IRS observations.

The morpho-kinematics of \Hmol\ present similarities with the CO filaments forming the CND in Cen~A \citep[analyzed by][]{espada_disentangling_2017}, and at larger kpc scales with filaments of inflowing CO in NGC~1275 \citep{salome_1275_2011,lim_2008}.
These filamentary structures are morphologically consistent with simulations of magnetized accreting molecular gas \citep[M87$^\star$,][]{guo_2024}, which also predict secondary magnetically driven polar outflows of molecular gas, although less significant. 
The \Hmol\ low dispersion spiral structure and the radial excitation gradient (see Fig.~\ref{fig:s5m2} and \ref{fig:M0}) point to a similar scenario.

The morpho-kinematics of the ionized gas are analyzed in details in the companion paper \citet{alonso-herrero_miconic_2025} and point to an outflow in the form of an expanding bubble driven by the jet, similar to those seen in AGN jet simulations \citep{mukherjee_2025}. Although a molecular outflow is detected at larger scales \citep[> 15" or 280~pc from the nucleus][]{israel_outflow_2017}, we only detect slight deviations in the wings of the \Hmol\ lines in the inner disk (see Sect.~\ref{subsec:H2kinematics} and Fig.~\ref{fig:disp_profiles}).
This pattern may trace enhanced velocity dispersion resulting from the adiabatic compression of infalling gas onto the central regions of the CND \citep{vollmer_quenching_2013}. Alternatively, it could reflect the presence of a radially outflowing component \citep{alonso-herrero_miconic_2025}, possibly arising from inhomogeneous emission within an expanding shell or bubble, where projection effects lead to asymmetric contributions from the approaching and receding sides. 

Both types of velocity structures are commonly seen in numerical simulations of CNDs \citep{guo_2024} and in jet-driven outflows \citep[e.g.,][]{mukherjee_2018,mukherjee_2025}. However without a more detailed modeling of the velocity fields and line profiles, it is not possible at this stage to distinguish between inflows and outflows in the \Hmol\ kinematics.

Overall, \Hmol\ lines exhibit velocity dispersions about one order of magnitude lower compared to the ionized gas. The Gaussian-fit sigma maps from \citet{alonso-herrero_miconic_2025} reveal that the velocity dispersion for the ionized gas is maximal in the ND, within $\sim1.5''$ of the AGN, and decreases along the jet. 
This net different behavior between the two phases has already been reported for other sources \citep[i.e. NGC~1275,][]{riffel_2020} and is thought to be caused by different gas density and dynamical coupling to the jet \citep{morganti_fast_2015, mukherjee_relativistic_2016, speranza_warm_2022}.

\section{Conclusions}

We have obtained pc-scale resolution mid-infrared spectral maps of the 100-200 pc inner-most region of Cen~A with the JWST/MRS, covering the 5-28~\mum\ wavelength range. This work, along with two companion papers on Cen~A \citep{alonso-herrero_miconic_2025,pantoni_2026} are part of the MIRI GTO program MICONIC. We focused here on spatially-resolved morpho-kinematics and excitation of the \Hmol\  rotational line emission. Our main results are the following:
\begin{itemize}[leftmargin=*, itemsep=0.2em]
    \item The \Hmol\ surface brightness maps show an inhomogeneous morphology with filaments of molecular gas peaking in the ICND at lower excitation and near the AGN at stronger excitation; lower levels also reveal a central cavity (40~pc in diameter).
    \item Velocity maps show global rotation plus a warped-disk geometry and non-circular motions. The filaments display lower dispersion, consistent with a coherent inflow toward the AGN, while residuals from Gaussian fits suggest a possible outflow at the base \citep[see][]{alonso-herrero_miconic_2025}. The molecular gas exhibits different morpho-kinematics from the ionized component, which peaks near the AGN and along the jet, showing higher dispersion and outflow signatures.
    \item Excitation diagrams indicate that the average temperatures of the bulk of the \Hmol\ are almost $\sim$2 times higher in the nucleus (416$\pm$12)~K than in the ICND (286$\pm$7)~K. The physical parameter maps show temperatures up to 340~K on the eastern ICND side, where the jet is pointed towards us. The filaments contain most of the molecular mass and show the lowest ortho-to-para ratio ($\sim1.8$).
    \item The mass of the molecular gas  obtained via fit of the excitation diagram at 100~K is $(9.6\pm4)\times10^5$~\Msun\ in the inner disk, while the dynamical mass is $(5\pm0.4)\times10^8$~M$_\odot$ within 4" (74~pc) of the AGN.

    \item The $L_\mathrm{H_2}/L_\mathrm{17\mu m}$ ratio increases by 2 orders of magnitude from the ND to the ICND and exceeds the threshold of photoionization processes at 1.3" (30~pc) from the AGN, indicating collisional excitation in the ICND. The $\Hmol / \mathrm{PAH}_{7.7\mu m}$ luminosity ratio exceeds the PDR model threshold for significant non-radiative excitation in all regions.
    \item The S(5) velocity dispersion at the PSF scale (6~pc) yields heating rates consistent with the dissipation rate $L_\mathrm{H_2}/M_\mathrm{H_2}=2.7~\mathrm{L_\odot}/\mathrm{M_\odot}$ measured along the coherent spiral streamer, supporting that dissipation of mechanical energy contributes to the excitation of \Hmol.
    \item The ratio $L_\mathrm{H_2}/L_\mathrm{X(2-10~keV)}$ exceeds by a factor 4 the upper limit predicted by XDR models, again supporting the importance of shocks in  heating the \Hmol\ gas. 
 \end{itemize}
The last three points in particular, are further evidence to the importance of shocks in heating \Hmol\ and in causing loss of angular momentum in the CND, as already suggested in existing literature \citep{espada_disentangling_2017, Ogle_2010}.
A forthcoming analysis presented in paper II will combine radiative and mechanical excitation models, based on the methodology presented in \cite{villa-velez_radiative_2024}, to quantify, on a spaxel-by-spaxel basis, the relative contributions of shocks, UV, and X-ray pumping to \Hmol\ excitation, disentangling the gas energy budget of the ICND.

\begin{acknowledgements}
    This work is based, in part, on observations made with the NASA/ESA/CSA James Webb Space Telescope. PG acknowledges  the Sorbonne University (FSI), the Centre National d'Etudes Spatiales (CNES), the `Programme National de Cosmologie and Galaxies' (PNCG) and the `Physique Chimie du Milieu Interstellaire' (PCMI) programs of CNRS/INSU, with INC/INP co-funded by CEA and CNES,  for there financial supports. AAH and LHM acknowledge support from grant PID2021-124665NB-I00 funded by MCIN/AEI/10.13039/501100011033 and by ERDF A way of making Europe. LP and MB acknowledge funding from the Belgian Science Policy Office (BELSPO) through the PRODEX project “JWST/MIRI Science exploitation” (C4000142239). RAR  acknowledges the support from the Conselho Nacional de Desenvolvimento Científico e Tecnológico (CNPq; Projects 303450/2022-3, and 403398/2023-1), the Coordenação de Aperfeiçoamento de Pessoal de Nível Superior (CAPES; Project 88887.894973/2023-00), and Fundação de Amparo à Pesquisa do Estado do Rio Grande do Sul (FAPERGS; Project 25/2551-0002765-9)
\end{acknowledgements}

\bibliographystyle{aa}  
\bibliography{references-1}

\appendix

\section{MRS spatial and spectral resolutions} \label{appendix:mrs}

We summarize here the technical specifications for each MRS channel and sub-channel. Table~\ref{tab:miri} details the FoV opening and centering, as well as the spatial and spectral resolutions and samplings. 
To estimate the spatial resolution as the FWHM of the PSF at the observation wavelength, we use the PSF characterization equation provided by \citet{law_3d_2023}:
\begin{equation}
\mathrm{FWHM}_{\mathrm{PSF}}\left[ \mathrm{arcsec} \right]=0.033'' \times\lambda\left[\mu\mathrm{m}\right]\,+0.106''
\end{equation} 
To estimate the spectral resolution power at different wavelengths we use the characterization equation provided by \citet{pontoppidan_2024}:
\begin{equation}
R(\lambda) = \frac{\lambda}{\Delta \lambda} = K_1 \ + \ K_2\ \times \ \lambda[\mathrm{\mu m}]
\label{eq:resolve}
\end{equation} 
where the factors $K_1$ and $K_2$ depend on the channels and the sub-channels and are reported in Table~\ref{tab:ab}.

\begin{table}[h]
   \caption{Coefficients for Eq.~\ref{eq:resolve} for the resolution power, by channel and sub-channel.}
\centering
\begin{tabular}{c c c}
\hline 
Subchannel & K$_1$ & K$_2$\\ \hline\hline
4C & -3601 & 216 \\ 
4B & -1176 & 150 \\  
4A & -2166 & 225 \\  
3C   & -2445 & 312 \\  
3B & -1871 & 317 \\  
3A  & -5120 & 633 \\ 
2C & -231 & 264 \\ 
2B & -331 & 400 \\ 
2A  & 332 & 400 \\ 
1C   & -543 & 601 \\ 
1B & 2742 & 150 \\  
1A  & -19.5 & 572 \\ 
\hline

\end{tabular}

   \tablefoot{\citet{pontoppidan_2024}.}
\label{tab:ab}

\end{table}

\begin{table*}[h]
\caption{MRS technical specs.}
\centering

\begin{tabular}{c c c c c c c c c c c}

\hline 
Channel & FoV & Center (RA, Dec)  & \multicolumn{2}{c}{PSF FWHM\textsuperscript{a}} & R\textsuperscript{b} & $\Delta V$ & \multicolumn{2}{c}{Spaxel size} & \multicolumn{2}{c}{$\delta \lambda$}\\
 & [arcsec$^2$] & [J2000] & ["] & [spaxel] &  & [\kms] & ["] & [parsec] & [nm] & [\kms]\\ 
\hline\hline

4-long   & $7.7\times 12$ & 13:25:27.688 -43:01:08.718 & 0.98 & 2.81 & 2134 & 140 & 0.35 & 6.45 & 6.0 & 68 \\ 
4-medium & $7.7\times 12$ & 13:25:27.688 -43:01:08.718 & 0.85 & 2.43 & 2206 & 136 & 0.35 & 6.45 & 6.0 & 80\\ 
4-short  & $7.7\times 12$ & 13:25:27.688 -43:01:08.718 & 0.74 & 2.11 & 2154 & 139 & 0.35 & 6.45 & 6.0 & 94\\ 

3-long   & $6.0 \times 10$ & 13:25:27.569 -43:01:08.754 & 0.65 & 3.26 & 2720 & 110 & 0.2 & 3.68 & 2.5 & 45\\ 
3-medium & $6.0 \times 10$ & 13:25:27.569 -43:01:08.754 & 0.58 & 2.90 & 2686 & 112 & 0.2 & 3.68 & 2.5 & 52\\ 
3-short  & $6.0 \times 10$ & 13:25:27.569 -43:01:08.754 & 0.52 & 2.58 & 2758 & 109 & 0.2 & 3.68 & 2.5 & 60\\ 

2-long   & $4.5 \times 8.6$ & 13:25:27.633 -43:01:08.787 & 0.46 & 2.72 & 2615 & 115 & 0.17 & 3.13 & 1.3 & 36\\ 
2-medium & $4.5 \times 8.6$ & 13:25:27.633 -43:01:08.787 & 0.41 & 2.44 & 3405 & 88  & 0.17 & 3.13 & 1.3 & 42\\
2-short  & $4.5 \times 8.6$ & 13:25:27.633 -43:01:08.787 & 0.37 & 2.19 & 3568 & 84  & 0.17 & 3.13 & 1.3 & 48\\ 

1-long   & $3.6 \times 7.5$ & 13:25:27.624 -43:01:08.296 & 0.34 & 2.60 & 3676 & 82  & 0.13 & 2.39 & 0.8 & 34\\
1-medium & $3.6 \times 7.5$ & 13:25:27.624 -43:01:08.296 & 0.31 & 2.36 & 3656 & 82  & 0.13 & 2.39 & 0.8 & 39\\ 
1-short  & $3.6 \times 7.5$ & 13:25:27.624 -43:01:08.296 & 0.28 & 2.16 & 3000 & 100 & 0.13 & 2.39 & 0.8 & 45\\

\hline
\end{tabular}
   \tablefoot{ FoV indicates the angular coverage for every channel. Center indicates the RA-Dec coordinates of the FoV's center for each channel. PSF FWHM expresses the channel spatial resolution, in arcsecs and spaxel. $R$ is the resolving power, which is converted to velocity ($\Delta V$) at the central wavelength of the subchannel. The spaxel size is provided in both arcsecs and parsecs (assuming a distance from Cen~A of 3.8~Mpc). $\delta \lambda$ is the spectral sampling step, expressed as both wavelength and velocity. }
\label{tab:miri}
 
\end{table*}

\section{Continuum measurements}\label{appendix:cont_tab}
We present in Table~\ref{table:continuum} the results of the 2D-Gaussian fit of the centroid of the continuum emission (see Sect.~\ref{subsec:continuum}), along with photometric measurements of the continuum, over each channel's FoV and in the ND.

\begin{table*}[h]
\caption{Results of the 2D-gaussian fit of the continuum maps and measurements of the continuum fluxes close to the \Hmol\ lines.}
\centering

\begin{center}
\begin{tabular}{c c c cc cc c c}
\hline 
$\lambda$ & $\Delta \lambda$ & Masked line  
& \multicolumn{2}{c}{PSF FWHM} 
& \multicolumn{2}{c}{Fitted FWHM} 
& Continuum flux (full FoV) & Continuum flux (ND)\\

[\mum] & [nm] & 0--0 
& ["] & [pc] 
& ["] & [pc] 
& [10\(^{-16}\) W~m$^{-2}$] 
& [10\(^{-16}\) W~m$^{-2}$] \\ 
\hline\hline

17.07 & 32.50 & S(1) & 0.67 & 12.34 & 0.71 & 13.08 & $8.44 \pm 0.02$ & $7.38 \pm 0.03$ \\ \hline
12.30 & 22.50 & S(2) & 0.51 & 9.39  & 0.59 & 10.87 & $8.75 \pm 0.03$ & $6.95 \pm 0.03$ \\ \hline
9.68  & 16.90 & S(3) & 0.43 & 7.92  & 0.43 & 7.92  & $1.47 \pm 0.04$ & $1.12 \pm 0.01$ \\ \hline
8.04  & 15.60 & S(4) & 0.37 & 6.82  & 0.41 & 7.55  & $3.63 \pm 0.04$ & $2.74 \pm 0.02$ \\ \hline
6.92  & 12.00 & S(5) & 0.33 & 6.08  & 0.35 & 6.45  & $1.85 \pm 0.05$ & $1.52 \pm 0.03$ \\ \hline
6.12  & 9.60  & S(6) & 0.31 & 5.71  & 0.34 & 6.26  & $1.90 \pm 0.06$ & $1.50 \pm 0.02$ \\ \hline
5.52  & 9.60  & S(7) & 0.29 & 5.34  & 0.32 & 5.89  & $2.04 \pm 0.08$ & $1.64 \pm 0.03$ \\ \hline
5.06  & 8.80  & S(8) & 0.27 & 4.97  & 0.31 & 5.71  & $2.04 \pm 0.08$ & $1.63 \pm 0.03$ \\ \hline

\end{tabular}
\end{center}

\tablefoot{ The spatial resolution (PSF FWHM in arcsec and parsec) is taken from \citet{law_3d_2023}. To allow comparison between the continuum and the \Hmol\ line fluxes, the continuum is computed close to the spectral line over the full FoV and within a circle of radius 2$\times$FWHM around the AGN for the ND. After masking the line, this flux is spectrally integrated around the central wavelength $\lambda$ of the \Hmol\  line, and over a wavelength range $\Delta \lambda$.} 
\label{table:continuum}

\hspace{1cm}\\
{\raggedright \textsuperscript{a}\small{\citet{law_3d_2023}} \\ \textsuperscript{b}\small{\citet{pontoppidan_2024}} \\%
}
\end{table*}

\section{Spectral extractions and line fluxes}
\label{appendix:spectra}
\begin{figure*}[ht]

    \includegraphics[width=.45\linewidth]{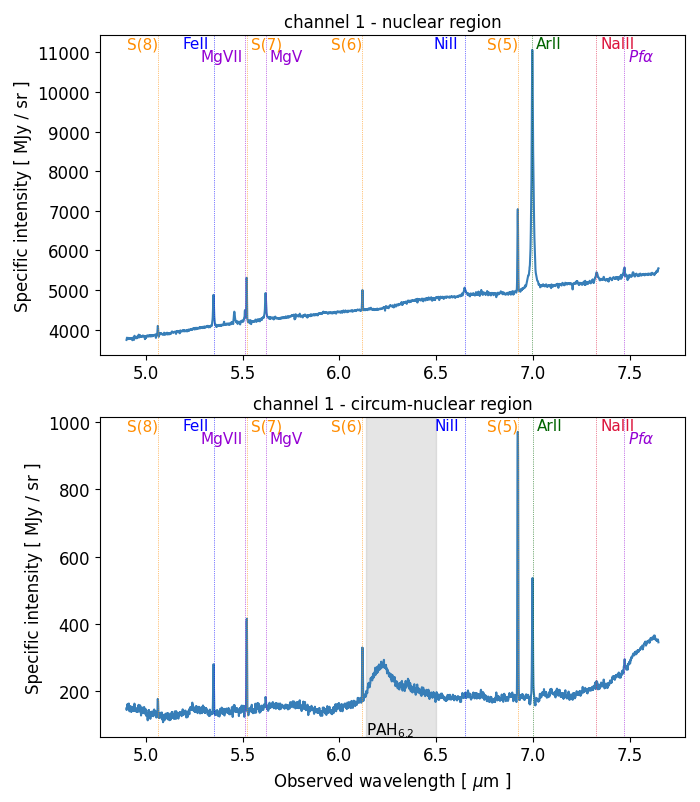}
    \hfill
    \includegraphics[width=.45\linewidth]{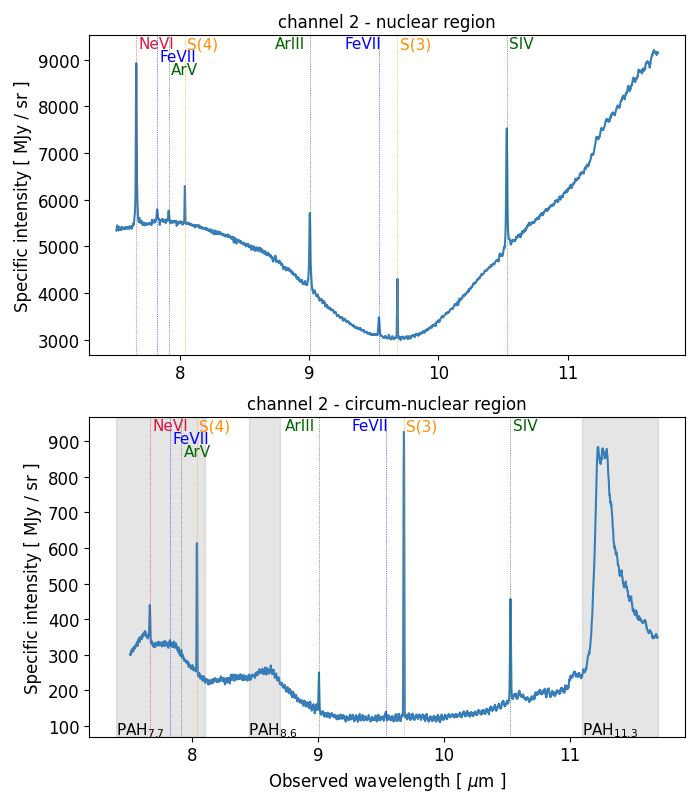}
    \includegraphics[width=.45\linewidth]{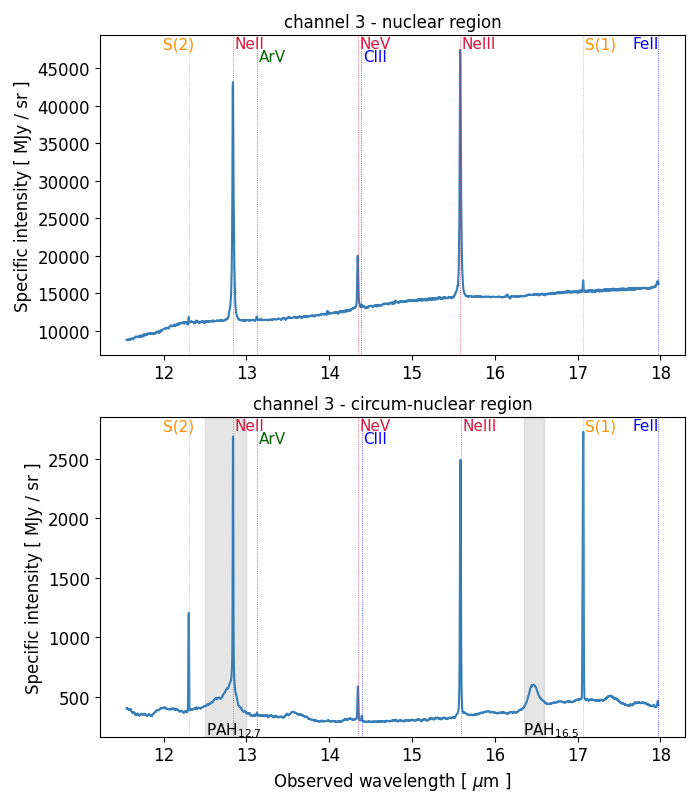}
    \hfill
    \includegraphics[width=.45\linewidth]{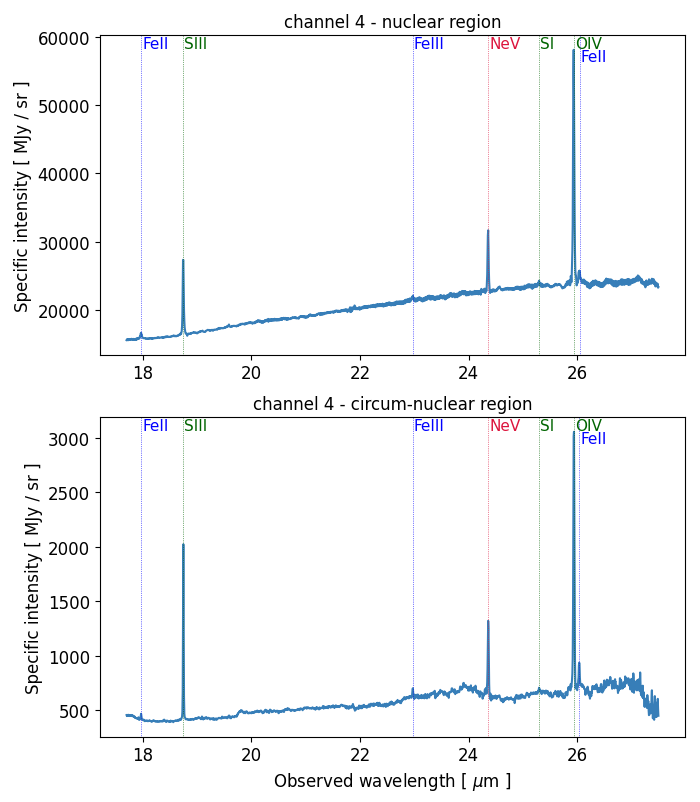}
    \caption{Spectrum of the inner region of Centaurus A. Zooms on the different MRS channels comparing the ND with the ICND, respectively inside and outside a 1.3"-radius circle (24 pc), as shown on the small inset image (continuum map at 17 µm) of Fig~\ref{fig:spectrum}. This aperture corresponds to 2×FWHM of the PSF at the 0–0 S(1) line. }
    \label{fig:spectra}
\end{figure*}

\begin{figure*}[htp]
    \centering
    \includegraphics[width=0.85\linewidth]{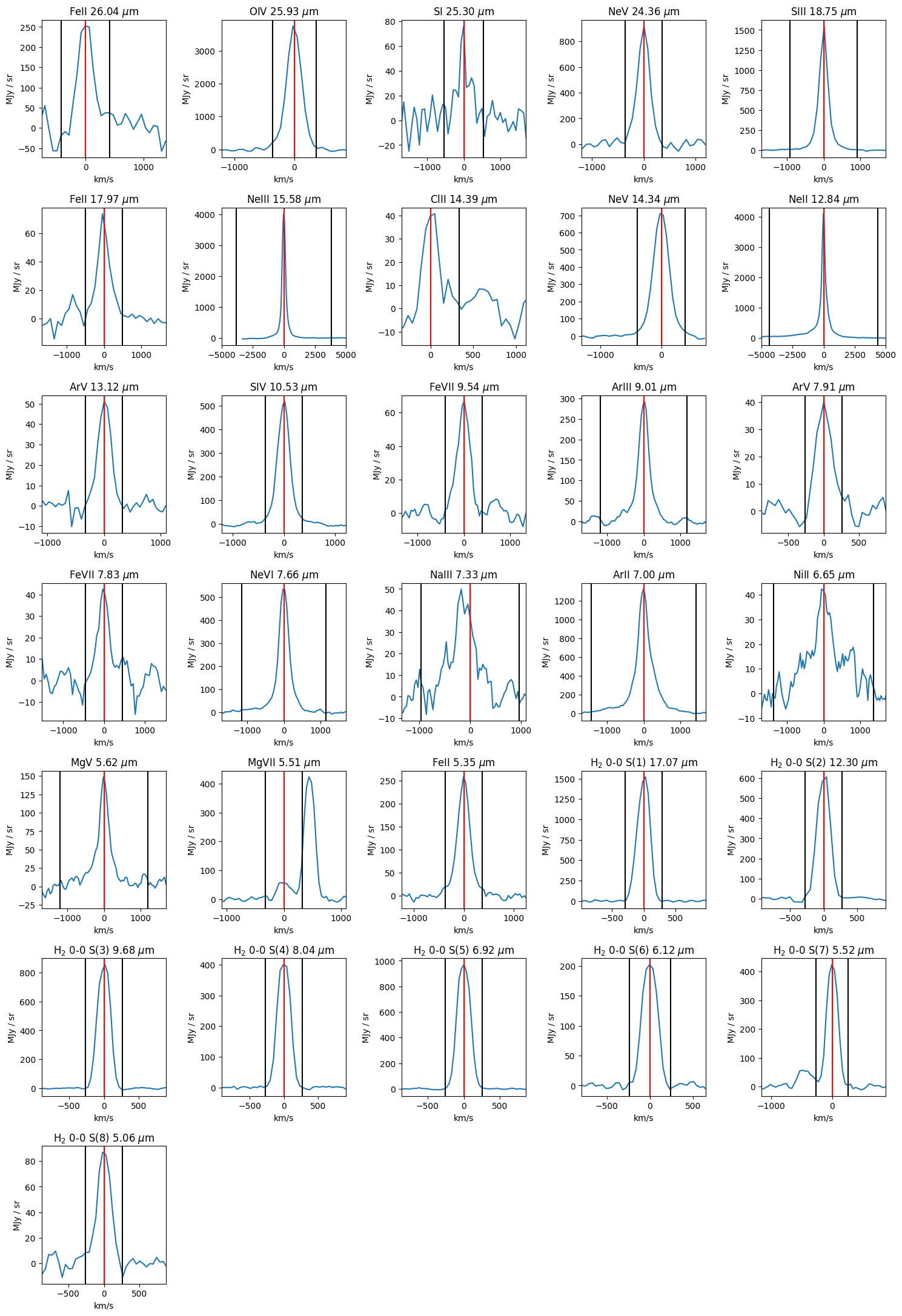}
    \caption{Continuum-subtracted line profiles averaged over the respective channel's full FoV. The black vertical bars represent the flux integration boundaries for each line. The [SIII], [NeIII], [NeII], [ArII], [ArIII], [NeVI], [MgV] are assumed to have Lorentzian profiles, and hence are wider  compared to the other lines. Notice that [ClII], [NaIII], and [NiII] present a strongly noisy continuum, which renders the fit of the profile challenging. }
    \label{fig:lineprof}
\end{figure*}

We present in Figure~\ref{fig:spectra} the ND and the ICND spectra of Cen~A from the four MRS channels. Extractions are performed respectively inside and outside a 1.3"-radius (24~pc) circle centered on the AGN. This aperture corresponds to 2$\times$FWHM of the PSF at 17~\mum. The extraction regions are displayed in Figure~\ref{fig:spectrum}.
In Figure~\ref{fig:lineprof} we present the line profiles extracted by averaging over the full FoV of the continuum subtracted sub-cubes. The black vertical lines mark the integration boundaries set to calculate the line fluxes listed in Table~\ref{tab:flux} (see Sect.~\ref{sec:inventory}).

\begin{sidewaystable*}[h]
\caption{Line fluxes extracted over different areas.}
\centering
\begin{tabular}{c c c c c c c c c c}
\hline
Line & $\lambda_\mathrm{obs}$ & \multicolumn{2}{c}{$2\times$FWHM} & \multicolumn{6}{c}{Flux}  \\
 & [$\mu$m] & ["] & [pc] & \multicolumn{6}{c}{[10$^{-17}$ W~m$^{-2}$]} \\
\cline{5-10}
 & & & & Full & ND & ICND & CH1 &  (a) & (b) \\
\hline
\hline

$[\mathrm{FeII}]$ & 26.04 & 1.93 & 35.51 & 7.24 $\pm$ 0.39 & 3.80 $\pm$ 0.31 & 3.43 $\pm$ 0.29 & 4.52 $\pm$ 0.39 & 0.08 $\pm$ 0.04 & 0.04 $\pm$ 0.04 \\
$[\mathrm{OIV}]$ & 25.93 & 1.92 & 35.33 & 101.51 $\pm$ 0.43 & 75.03 $\pm$ 0.42 & 26.48 $\pm$ 0.24 & 78.32 $\pm$ 0.43 & 1.02 $\pm$ 0.04 & 2.79 $\pm$ 0.04 \\
$[\mathrm{SI}]$ & 25.30 & 1.88 & 34.59 & 2.13 $\pm$ 0.29 & 1.50 $\pm$ 0.22 & 0.63 $\pm$ 0.22 & 1.66 $\pm$ 0.29 & 0.04 $\pm$ 0.03 & 0.00 $\pm$ 0.03 \\
$[\mathrm{NeV}]$ & 24.36 & 1.82 & 33.49 & 24.76 $\pm$ 0.34 & 18.98 $\pm$ 0.31 & 5.79 $\pm$ 0.20 & 20.04 $\pm$ 0.34 & 0.13 $\pm$ 0.03 & 0.34 $\pm$ 0.02 \\
$[\mathrm{SIII}]$ & 18.75 & 1.45 & 26.68 & 59.88 $\pm$ 0.50 & 41.64 $\pm$ 0.53 & 18.24 $\pm$ 0.15 & 43.78 $\pm$ 0.50 & 0.56 $\pm$ 0.02 & 1.08 $\pm$ 0.02 \\
$[\mathrm{FeII}]$ & 17.97 & 1.40 & 25.76 & 3.59 $\pm$ 0.20 & 2.88 $\pm$ 0.21 & 0.71 $\pm$ 0.07 & 3.23 $\pm$ 0.20 & 0.03 $\pm$ 0.01 & 0.07 $\pm$ 0.01 \\
$[\mathrm{NeIII}]$ & 15.58 & 1.24 & 22.82 & 156.44 $\pm$ 0.45 & 121.01 $\pm$ 0.69 & 35.43 $\pm$ 0.23 & 144.96 $\pm$ 0.45 & 1.25 $\pm$ 0.04 & 2.92 $\pm$ 0.04 \\
$[\mathrm{NeV}]$ & 14.34 & 1.16 & 21.34 & 25.41 $\pm$ 0.04 & 19.52 $\pm$ 0.04 & 5.89 $\pm$ 0.03 & 23.89 $\pm$ 0.04 & 0.10 $\pm$ 0.01 & 0.34 $\pm$ 0.01 \\
$[\mathrm{ClII}]$ & 14.39 & 1.16 & 21.34 & 0.90 $\pm$ 0.03 & 0.22 $\pm$ 0.03 & 0.68 $\pm$ 0.02 & 0.59 $\pm$ 0.03 & 0.05 $\pm$ 0.01 & 0.04 $\pm$ 0.01 \\
$[\mathrm{NeII}]$ & 12.84 & 1.06 & 19.50 & 283.30 $\pm$ 0.75 & 178.27 $\pm$ 0.77 & 105.03 $\pm$ 0.56 & 236.81 $\pm$ 0.75 & 4.03 $\pm$ 0.10 & 7.00 $\pm$ 0.13 \\
$[\mathrm{ArV}]$ & 13.12 & 1.08 & 19.87 & 1.69 $\pm$ 0.04 & 1.29 $\pm$ 0.05 & 0.40 $\pm$ 0.03 & 1.56 $\pm$ 0.04 & -- & 0.02 $\pm$ 0.01 \\
$[\mathrm{SIV}]$ & 10.53 & 0.91 & 16.74 & 14.51 $\pm$ 0.04 & 7.87 $\pm$ 0.03 & 6.64 $\pm$ 0.03 & 13.55 $\pm$ 0.04 & 0.12 $\pm$ 0.01 & 0.26 $\pm$ 0.01 \\
$[\mathrm{FeVII}]$ & 9.54 & 0.84 & 15.46 & 2.27 $\pm$ 0.06 & 1.46 $\pm$ 0.03 & 0.80 $\pm$ 0.05 & 2.22 $\pm$ 0.06 & 0.00 $\pm$ 0.01 & 0.04 $\pm$ 0.01 \\
$[\mathrm{ArIII}]$ & 9.01 & 0.81 & 14.90 & 12.45 $\pm$ 0.16 & 8.68 $\pm$ 0.09 & 3.77 $\pm$ 0.14 & 12.30 $\pm$ 0.16 & 0.04 $\pm$ 0.04 & 0.16 $\pm$ 0.03 \\
$[\mathrm{ArV}]$ & 7.91 & 0.73 & 13.43 & 1.30 $\pm$ 0.04 & 0.82 $\pm$ 0.02 & 0.48 $\pm$ 0.04 & 1.24 $\pm$ 0.04 & 0.00 $\pm$ 0.01 & 0.01 $\pm$ 0.01 \\
$[\mathrm{FeVII}]$ & 7.83 & 0.73 & 13.43 & 1.93 $\pm$ 0.09 & 1.33 $\pm$ 0.05 & 0.60 $\pm$ 0.07 & 1.89 $\pm$ 0.09 & 0.01 $\pm$ 0.02 & 0.02 $\pm$ 0.02 \\
$[\mathrm{NeVI}]$ & 7.66 & 0.72 & 13.25 & 26.85 $\pm$ 0.42 & 18.64 $\pm$ 0.38 & 8.21 $\pm$ 0.21 & 26.71 $\pm$ 0.42 & 0.01 $\pm$ 0.04 & 0.19 $\pm$ 0.04 \\
$[\mathrm{NaIII}]$ & 7.33 & 0.70 & 12.88 & 2.38 $\pm$ 0.20 & 1.63 $\pm$ 0.14 & 0.76 $\pm$ 0.15 & 2.38 $\pm$ 0.20 & 0.12 $\pm$ 0.04 & 0.12 $\pm$ 0.05 \\
$[\mathrm{ArII}]$ & 7.00 & 0.67 & 12.33 & 66.71 $\pm$ 0.31 & 55.43 $\pm$ 0.16 & 11.27 $\pm$ 0.26 & 66.71 $\pm$ 0.31 & 0.03 $\pm$ 0.07 & 0.13 $\pm$ 0.08 \\
$[\mathrm{NiII}]$ & 6.65 & 0.65 & 11.96 & 3.63 $\pm$ 0.25 & 2.25 $\pm$ 0.10 & 1.38 $\pm$ 0.22 & 3.63 $\pm$ 0.25 & 0.01 $\pm$ 0.06 & 0.03 $\pm$ 0.07 \\
$[\mathrm{MgV}]$ & 5.62 & 0.58 & 10.67 & 8.17 $\pm$ 0.51 & 5.12 $\pm$ 0.38 & 3.05 $\pm$ 0.35 & 8.17 $\pm$ 0.51 & 0.00 $\pm$ 0.10 & 0.09 $\pm$ 0.11 \\
$[\mathrm{MgVII}]$ & 5.51 & 0.58 & 10.67 & 3.04 $\pm$ 0.10 & 1.54 $\pm$ 0.04 & 1.50 $\pm$ 0.09 & 3.04 $\pm$ 0.10 & 0.00 $\pm$ 0.03 & 0.00 $\pm$ 0.03 \\

\vspace{1 mm}\\

$\mathrm{H_{2}\,0\!-\!0\ S(1)}$ & 17.07 & 1.34 & 24.66 & 33.29 $\pm$ 0.02 & 2.18 $\pm$ 0.03 & 31.11 $\pm$ 0.02 & 17.34 $\pm$ 0.02 & 1.79 $\pm$ 0.01 & 1.89 $\pm$ 0.01 \\
$\mathrm{H_{2}\,0\!-\!0\ S(2)}$ & 12.30 & 1.02 & 18.77 & 18.84 $\pm$ 0.03 & 1.04 $\pm$ 0.03 & 17.80 $\pm$ 0.02 & 11.05 $\pm$ 0.03 & 1.16 $\pm$ 0.01 & 1.14 $\pm$ 0.01 \\
$\mathrm{H_{2}\,0\!-\!0\ S(3)}$ & 9.68 & 0.85 & 15.64 & 19.62 $\pm$ 0.04 & 1.04 $\pm$ 0.01 & 18.58 $\pm$ 0.03 & 16.64 $\pm$ 0.04 & 1.53 $\pm$ 0.01 & 1.51 $\pm$ 0.01 \\
$\mathrm{H_{2}\,0\!-\!0\ S(4)}$ & 8.04 & 0.74 & 13.62 & 11.88 $\pm$ 0.04 & 0.76 $\pm$ 0.02 & 11.11 $\pm$ 0.04 & 10.49 $\pm$ 0.04 & 0.92 $\pm$ 0.01 & 0.88 $\pm$ 0.01 \\
$\mathrm{H_{2}\,0\!-\!0\ S(5)}$ & 6.92 & 0.67 & 12.33 & 22.28 $\pm$ 0.05 & 1.88 $\pm$ 0.03 & 20.40 $\pm$ 0.04 & 22.28 $\pm$ 0.05 & 1.75 $\pm$ 0.01 & 1.68 $\pm$ 0.01 \\
$\mathrm{H_{2}\,0\!-\!0\ S(6)}$ & 6.12 & 0.62 & 11.41 & 5.25 $\pm$ 0.06 & 0.47 $\pm$ 0.02 & 4.79 $\pm$ 0.05 & 5.26 $\pm$ 0.06 & 0.32 $\pm$ 0.01 & 0.35 $\pm$ 0.01 \\
$\mathrm{H_{2}\,0\!-\!0\ S(7)}$ & 5.52 & 0.58 & 10.67 & 12.37 $\pm$ 0.08 & 1.44 $\pm$ 0.03 & 10.93 $\pm$ 0.07 & 12.37 $\pm$ 0.08 & 0.83 $\pm$ 0.02 & 0.76 $\pm$ 0.02 \\
$\mathrm{H_{2}\,0\!-\!0\ S(8)}$ & 5.06 & 0.55 & 10.12 & 2.85 $\pm$ 0.08 & 0.29 $\pm$ 0.03 & 2.57 $\pm$ 0.07 & 2.85 $\pm$ 0.08 & 0.20 $\pm$ 0.02 & 0.18 $\pm$ 0.02 \\

\vspace{1 mm}\\
\hline

$\mathrm{Log_{10}}\ (\ L_{\mathrm{H_2}} \ ) $ [W] & -- & -- & -- 
& 32.34 & 31.20 & 32.31 & 32.23 & 31.18 & 31.18 \\
$\mathrm{Log_{10}}\ (\ L_{\mathrm{X(2-10\ keV)}} \ ) $ [W] & -- & -- & -- 
& 34.70 & 34.50 & 34.20 & -- & -- & -- \\

\hline

    \end{tabular}
    \tablefoot{Areas : the entire FoV of the respective channel (Full), the ND within an aperture of radius $2\times$FWHM (the extent of which is provided in the third column in arcsec and pc), outside the ND (ICND), and the area covered by the FoV of CH1. The line fluxes are computed via spectral profile integration, with boundaries displayed in Appendix~\ref{appendix:spectra} (Figure~\ref{fig:lineprof}), and are aperture corrected based on the correction factors from \citet{law_2025}. We also provide extractions from the Northern hotspot (a) and the Southern hotspot (b) (see Section~\ref{sect:morph} and Figure~\ref{fig:moment 0 maps}). For the [ArV] line, the noisy background makes it impossible to obtain a meaningful result in region (a). The last two lines provide respectively the logarithmic luminosities of the summed \Hmol\ lines and the X-ray luminosities obtained from \cite{Ogle_2010,kraft_2002} (both in Watt).}
    \label{tab:flux}
\end{sidewaystable*}

\section{Continuum and spectral maps}
\label{appendix:maps}
\begin{figure*}[htp]

        \includegraphics[width=\textwidth]{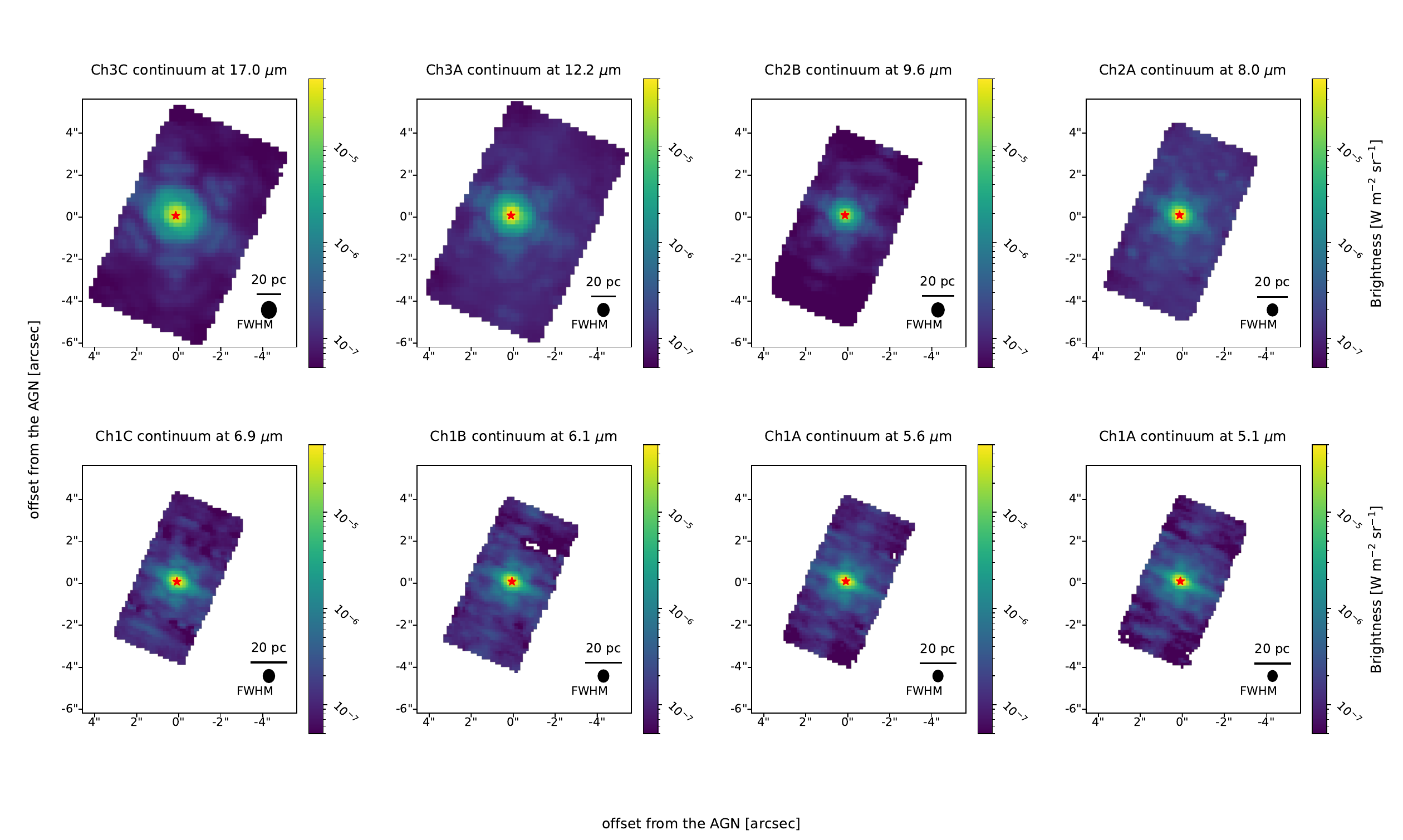}
    \caption{Continuum maps extracted around the \Hmol\ rotational lines (with the line masked). The red-star symbol marks the position of the central AGN: in RA-Dec 13:25:27.63 -43:01:08.30 (J2000).} 
    \label{fig:continuum}
    
\end{figure*}

\begin{figure*}[htp]
    \centering
    \includegraphics[width=\linewidth]{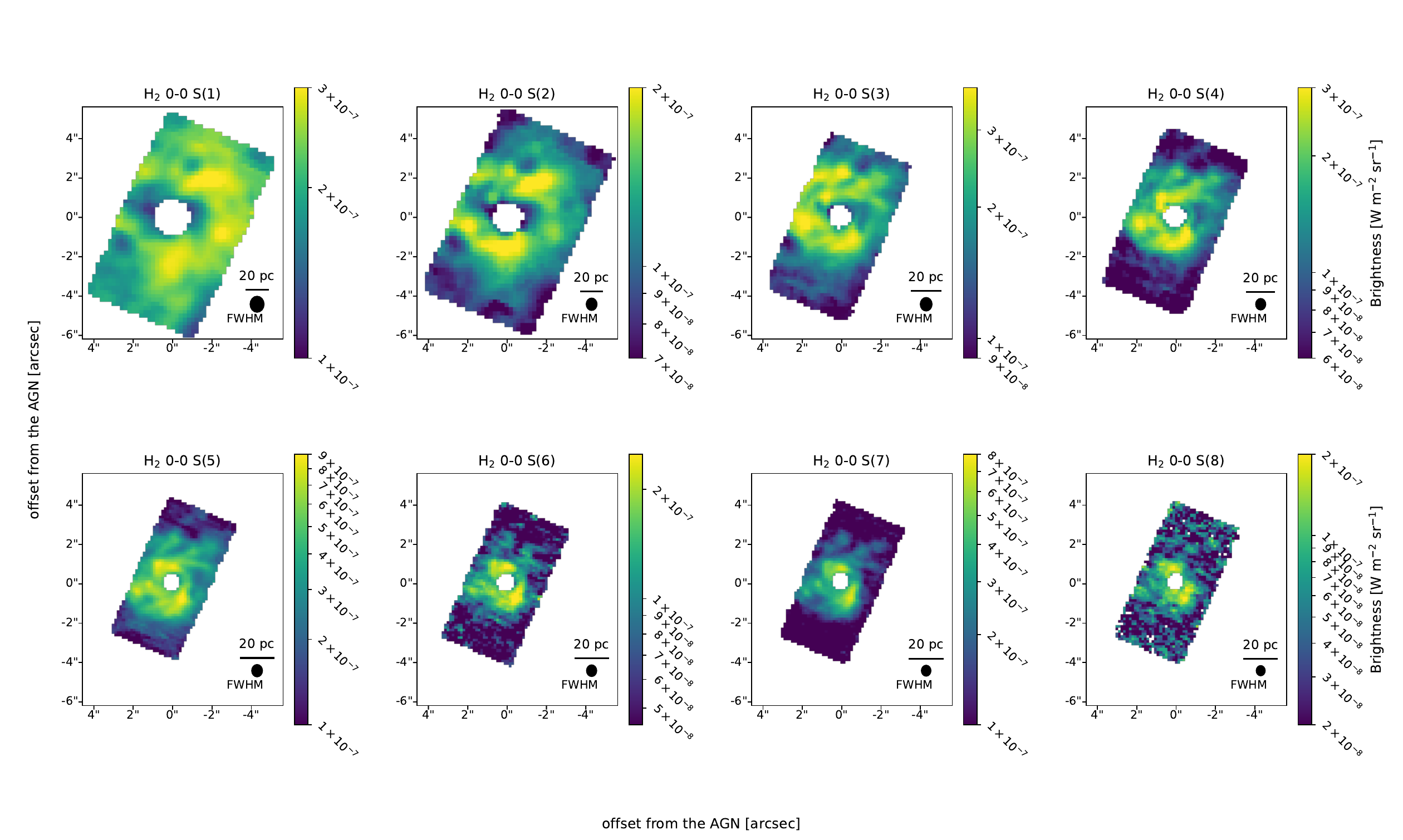}
    \caption{\Hmol\ rotational line surface brightness maps in [W m$^{-2}$ sr$^{-1}$] for all transitions from 0--0~S(1) to S(8). Central spaxels are masked. Spaxels with undetected lines are replaced with background noise level. The figures reveal how the brighter spots spiral towards the AGN along the \Hmol\ filaments (white contours in Fig \ref{fig:s5m2}), as excitation becomes stronger.} 

    \label{fig:M0}
\end{figure*}

\begin{figure*}[htp]
    \centering
    \includegraphics[width=\linewidth]{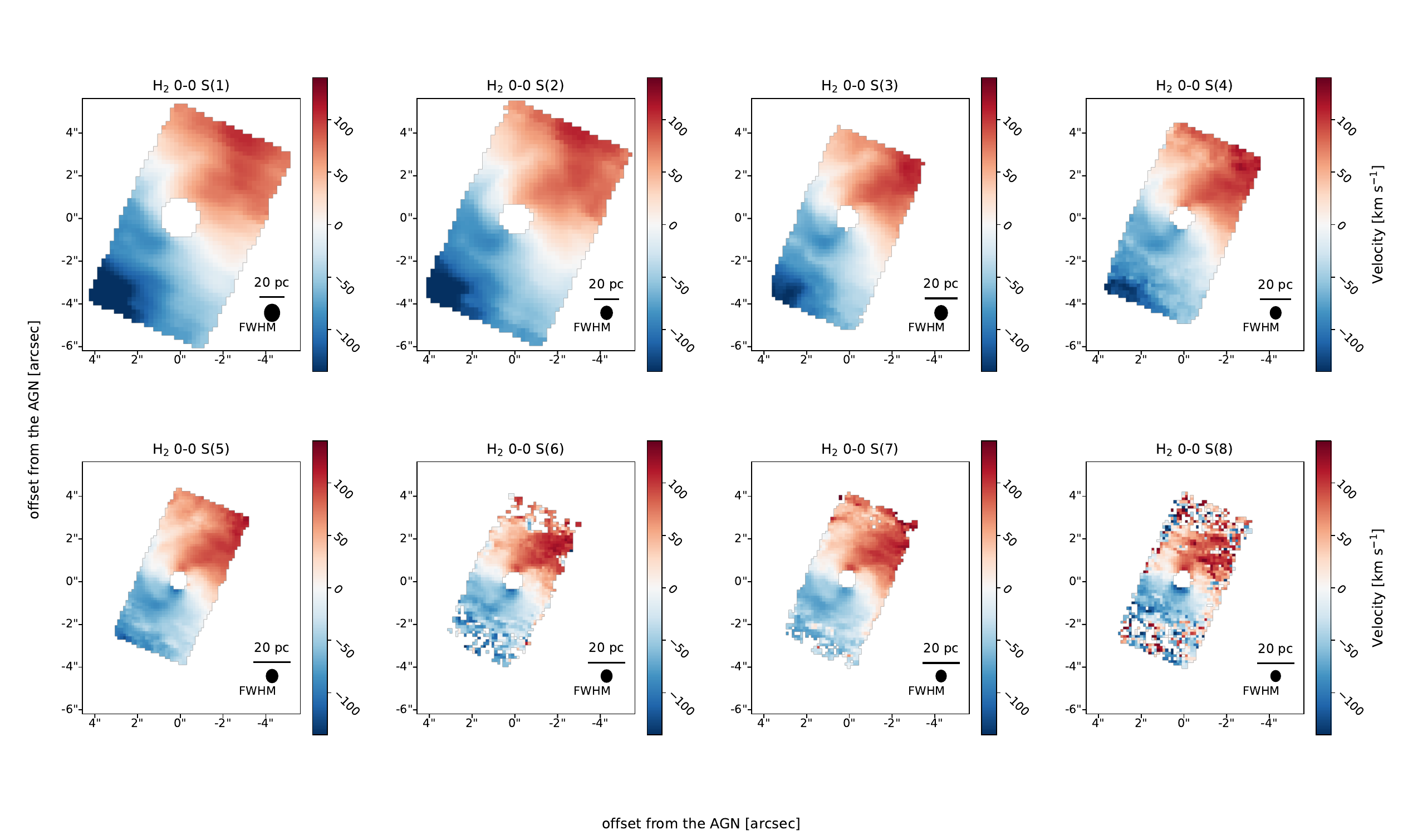}
    \caption{Velocity maps in [\kms]. Central spaxels and no-line-detection spaxels are masked. The maps show clear signs of global rotation, along with an evident S-shaped distortion due to the warped-disk geometry of the CND plus non-circular motion.}
    \label{fig:M1}
\end{figure*}

\begin{figure*}[htp]
    \centering
    \includegraphics[width=\linewidth]{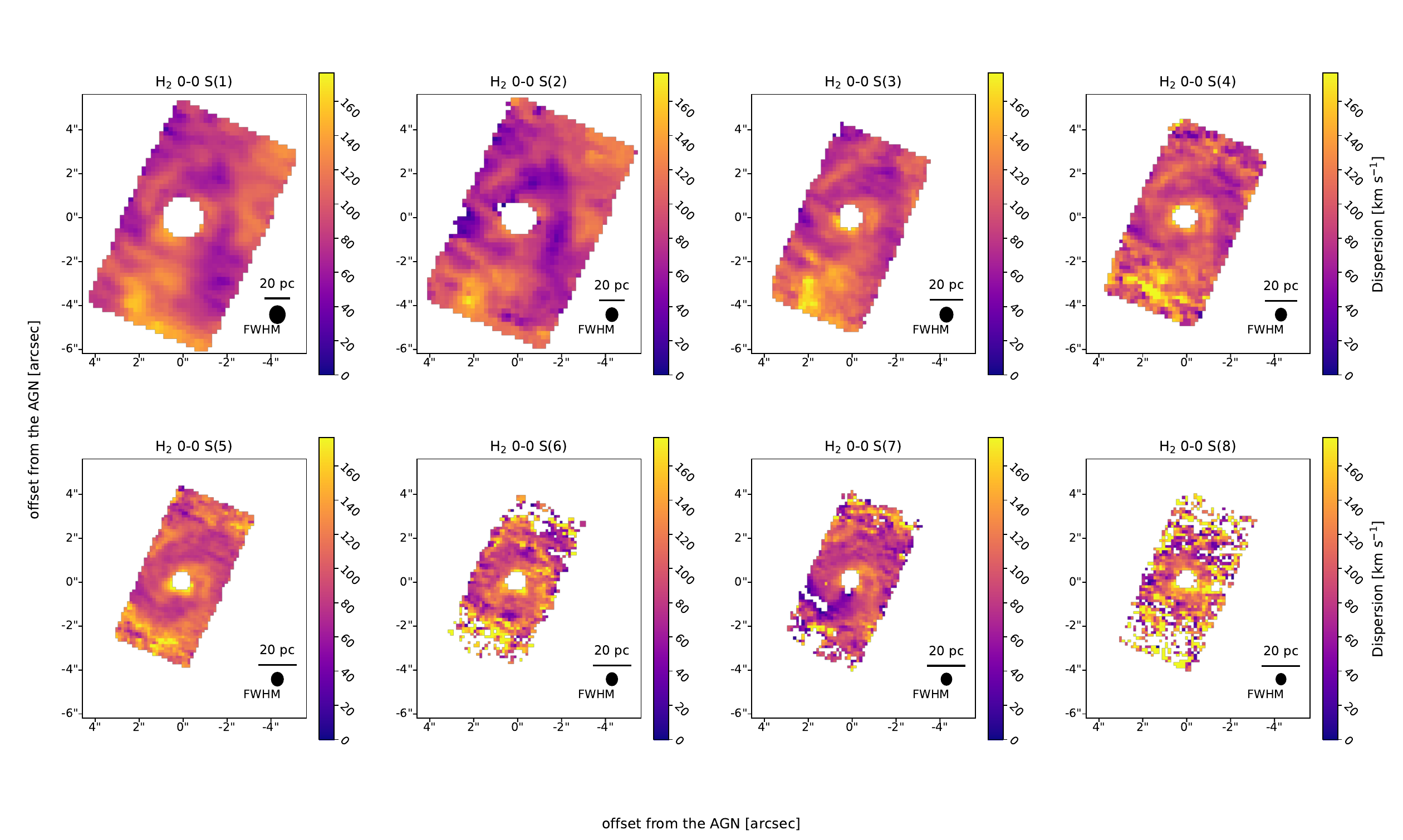}
    \caption{Velocity dispersion maps in [\kms]. Central spaxels and no-line-detection spaxels are masked. }
    \label{fig:M2}
\end{figure*}

We present here all the generated maps for \Hmol\ and for the continuum. Figure~\ref{fig:continuum} shows the maps of the continuum near each of the \Hmol\ lines. Figures~\ref{fig:M0}, \ref{fig:M1}, and \ref{fig:M2} show all the \Hmol\ maps respectively of moment-0, 1, and 2. Details are provided in the main text. For the \Hmol\ 0--0 S(8), we only show the moment maps generated with the $1\sigma_\mathrm{STD}$ spaxel flagging criterion since this method recovers spaxels in the ICND that reveal morphological correlations with the other maps.

\section{Power-law fits of the excitation diagrams}
\label{appendix:h2p}

We present here the power-law fits of the excitation diagrams performed using the H2powerlaw tool by \citet{togi_2016}, over the FoV of CH1, the ND, and the ICND. 
The tool fits the \Hmol\ lines assuming a continuous temperature distribution following the relation:
\begin{equation}
    \mathrm{d}N = \ m\ T^{-n} \ \mathrm{d}T
    \label{eq:togi}
\end{equation}
where d$N$ is the column density between $T$ and $T+\mathrm{d}T$, $n$ is the power-law index, and $m$ is a constant. Integrating \ref{eq:togi} between two temperature bounds $T_\mathrm{low}$ and $T_\mathrm{upp}$, and assuming that $T$ is equal to the rotational temperature, $m$ is found to be:
\begin{equation}
    m=\frac{N_\mathrm{tot}(n-1)}{T^{1-n}_\mathrm{low}-T^{1-n}_\mathrm{upp}}
\end{equation}
where $N_\mathrm{tot}$ is the total \Hmol\ column density in the FoV. Above 1000~K,  \cite{togi_2016} indicate that $T_\mathrm{upp}$ has a negligible impact on the mass estimate and it is hence fixed by default at 2000~K. The parameters $T_\mathrm{low}$ and $n$ are adjusted to fit the column densities from the excitation diagram, allowing to calculate $N_\mathrm{tot}$ and thus, given the mass of one \Hmol\ molecule $m_\mathrm{H_2}$, the FoV surface angle $\Omega$, and the distance from the source $d$, the total \Hmol\ mass is  $M_\mathrm{tot}=m_\mathrm{H_2} \ N_\mathrm{tot} \ \Omega \ d^2$.

\begin{figure*}[ht]

    \includegraphics[width=0.5\linewidth, trim=0 0 0 20, clip]{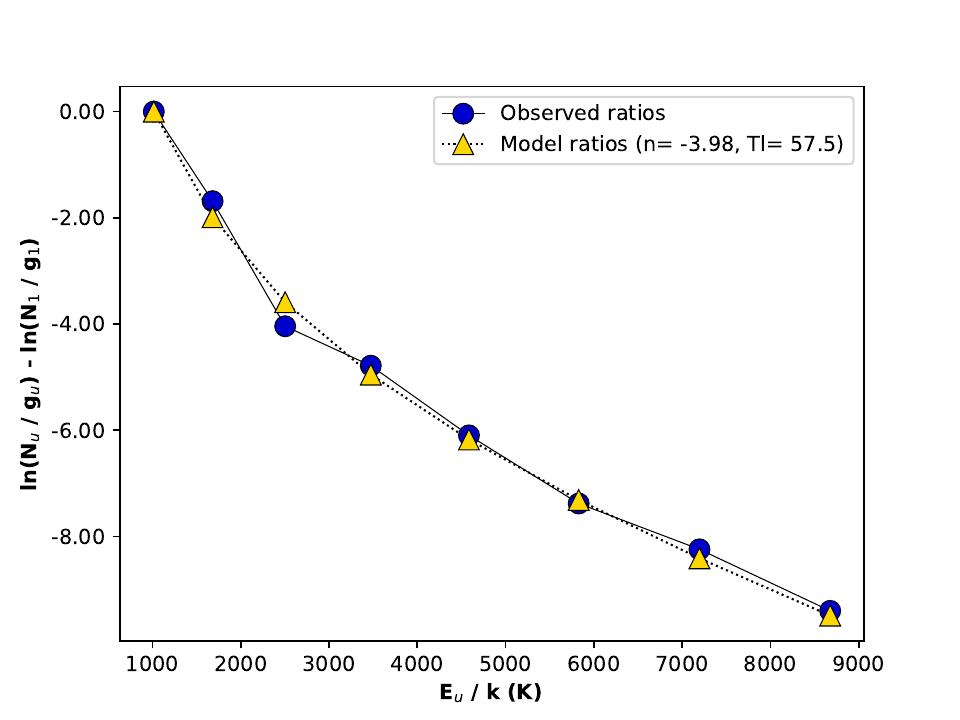}
    \put(-70, 130){\color{black}\text{CH1}} 
    \includegraphics[width=0.5\linewidth, trim=0 0 0 20, clip]{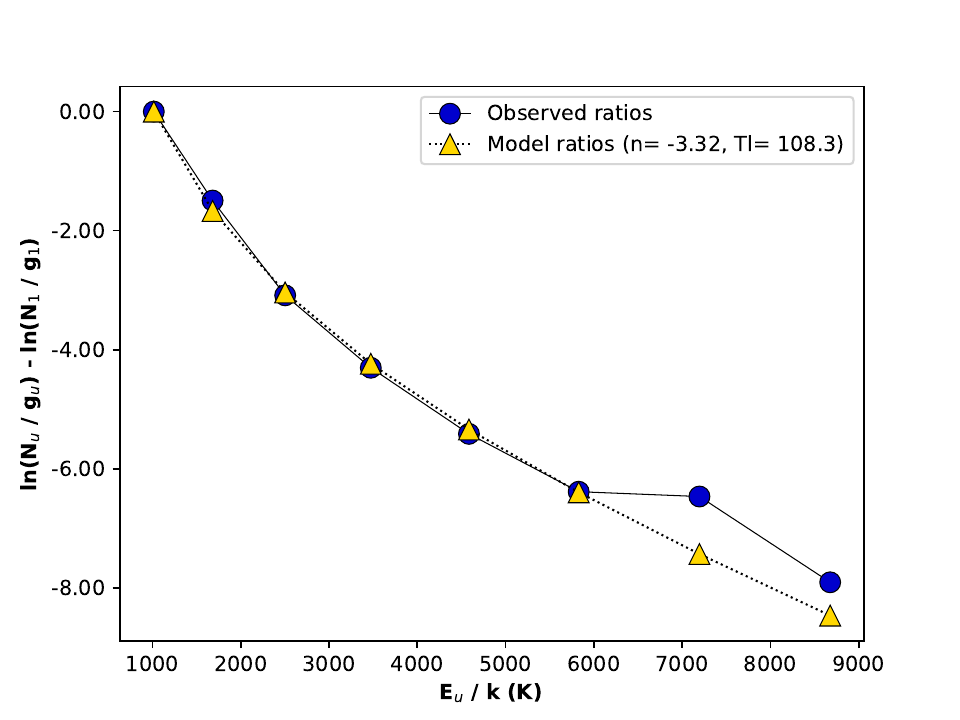}
    \put(-70, 130){\color{black}\text{ND}} \\
    \includegraphics[width=0.5\linewidth, trim=0 0 0 20, clip]{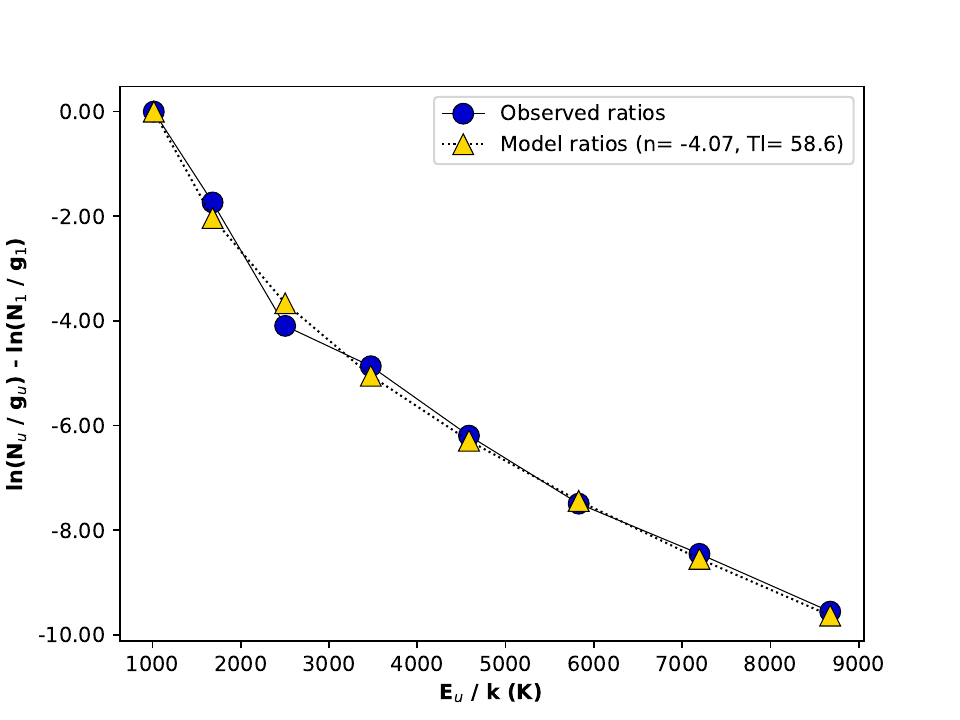}
    \put(-70, 130){\color{black}\text{ICND}} 
    \caption{Power-law fit of the excitation diagrams, performed using the H2Powerlaw tool from \citet{togi_2016} on the fluxes  extracted from the averaged cubes, over the full FoV of CH1, inside (ND), and outside (ICND) a circle centered around the AGN of radius 2$\times$FWHM. The legends report the results of the fit: \texttt{n} is the slope of the power-law, while \texttt{Tl} is the lower temperature limit of the fit $T_\mathrm{low}$.
    The S(3) line in the ND is corrected for extinction via the method presented by \citet{reefe_2025} (see App.~\ref{appendix:ext}). 
    The deviation of the S(7) from the fit in the ND is due to the difficult de-blending of the [MgVII] line at 5.51~\mum.}
    \label{fig:ex_pl}
\end{figure*}

Figure~\ref{fig:ex_pl} shows the results of the power-law fits on our \Hmol\ excitation diagrams, over the full FoV, the ND, and the ICND.

\section{Absorption in the silicate band}
\label{appendix:ext}

\cite{Ogle_2010} use a continuum fit to estimate absorption and find a non-negligible 0.4 relative extinction in the silicate band, near the \Hmol\ 0--0 S(3) line at 9.68~\mum. We present here the method we used to estimate the optical depth $\tau_\mathrm{9.7\mu m}$, based on \citet{reefe_2025}.

We mask the S(3) line in the excitation diagrams and perform a power-law fit (see App.~\ref{appendix:h2p}) to infer the intrinsic S(3) emission from the other lines. Flux ratios between the $i$-th and the $j$-th lines are expected to follow the relation:
\begin{equation}
    \frac{F_i}{F_j}=\frac{N_i\ A_i \ \lambda_j}{N_j\ A_j \ \lambda_i}
\end{equation}
were $F_i$, $N_i$, $A_i$, and $\lambda_i$ are respectively the flux, the column density, the Einstein coefficient, and the wavelenght of the $i$-th line.
The next \Hmol\ line adjacent S(3) is S(4) which is out of the silicate band. We hence compare the intrinsic flux ratio $[F_\mathrm{S(4)}/F_\mathrm{S(3)}]_\mathrm{intrinsic}$ inferred from the fit of the excitation diagram, with the observed ratio $[F_\mathrm{S(4)}/F_\mathrm{S(3)}]_\mathrm{observed}$. The optical depth at 9.68~\mum is therefore estimated as:
\begin{equation}
    \tau_\mathrm{9.7\mu m} = \mathrm{ln}\left(\frac{[F_\mathrm{S(4)}/F_\mathrm{S(3)}]_\mathrm{observed}}{[F_\mathrm{S(4)}/F_\mathrm{S(3)}]_\mathrm{intrinsic}}\right)
\end{equation}
We use this result to correct the S(3) line flux in our nuclear excitation diagram.

\end{document}